\documentclass{article}
 \usepackage{multicol}
 \usepackage{xcolor}
\usepackage{mdframed}
\usepackage{ragged2e}
 \usepackage{comment}
\usepackage{arxiv}
\usepackage{caption}
\DeclareCaptionLabelFormat{nospace}{#1#2}
\usepackage[utf8]{inputenc} 
\usepackage[T1]{fontenc}    
\usepackage{hyperref}       
\usepackage{url}            
\usepackage{booktabs}       
\usepackage{amsfonts}       
\usepackage{nicefrac}       
\usepackage{microtype}      
\usepackage{lipsum}		
\usepackage{graphicx}
\usepackage{doi}
\usepackage{enumitem}
\usepackage{subcaption}
\usepackage{tabularx}
\usepackage{longtable}
\title{I-SIRch: AI-powered Concept Annotation Tool for Equitable Extraction and Analysis of Safety Insights from Maternity Investigations}

\author{ \href{https://orcid.org/0000-0001-7736-5583}{\includegraphics[scale=0.06]{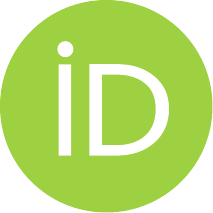}\hspace{1mm}Mohit Kumar~Singh {$\ddagger$}} \\
	Department of Computer Science\\
	School of Science, Loughborough University\\
	Loughborough, United Kingdom, LE11 3TU\\
	\And
	\href{https://orcid.org/0000-0002-4663-6907}{\includegraphics[scale=0.06]{orcid.pdf}\hspace{1mm}Georgina~Cosma{$\ddagger$}}\thanks{Corresponding Author: g.cosma@lboro.ac.uk; $\ddagger$ \texttt{These authors contributed equally to the work.}}\\
	Department of Computer Science\\
	School of Science, Loughborough University\\
	Loughborough, United Kingdom, LE11 3TU \\
 \And
	Patrick Waterson \\
	School of Design and Creative Arts\\
	Loughborough University\\
	Loughborough, United Kingdom, LE11 3TU\\
	\texttt{} \\
 \And
	Jonathan Back \\
	Health Services Safety Investigations Body (HSSIB)\\
	United Kingdom\\
	\texttt{} \\
 \And
	Gyuchan Thomas Jun \\
	School of Design and Creative Arts\\
	Loughborough University\\
	Loughborough, United Kingdom, LE11 3TU\\
	\texttt{} \\
}

\renewcommand{\shorttitle}{I-SIRch: AI-Powered Concept Annotation Tool}

\hypersetup{
pdftitle={ISRIch paper},
pdfsubject={cs.CL,cs.SE},
pdfauthor={Mohit K.~Singh, Georgina~Cosma, Patrick ~Waterson, Jonathan~Back, Gyuchan Thomas ~Jun},
pdfkeywords={I-SIRCH, Maternity research, Healthcare Safety Investigation},
}

\begin{document}
\maketitle
\begin{abstract}
Maternity care is a complex system involving treatments and interactions between patients, providers, and the care environment. To improve patient safety and outcomes, understanding the human factors (e.g. individuals decisions, local facilities) influencing healthcare delivery is crucial. However, most current tools for analysing healthcare data focus only on biomedical concepts  (e.g. health conditions, procedures and tests), overlooking the importance of human factors. 
We developed a new approach called I-SIRch, using artificial intelligence to automatically identify and label human factors concepts in maternity healthcare investigation reports describing adverse maternity incidents produced by England's Healthcare Safety Investigation Branch (HSIB). These incident investigation reports aim to identify opportunities for learning and improving maternal safety across the entire healthcare system. 
I-SIRch was trained using real data and tested on both real and simulated data to evaluate its performance in identifying human factors concepts. When applied to real reports, the model achieved a high level of accuracy, correctly identifying relevant concepts in 90\% of the sentences from 97 reports. Applying I-SIRch to analyse these reports revealed that certain human factors disproportionately affected mothers from different ethnic groups. 
Our work demonstrates the potential of using automated tools to identify human factors concepts in maternity incident investigation reports, rather than focusing solely on biomedical concepts. This approach opens up new possibilities for understanding the complex interplay between social, technical, and organisational factors influencing maternal safety and population health outcomes. By taking a more comprehensive view of maternal healthcare delivery, we can develop targeted interventions to address disparities and improve maternal outcomes.
\end{abstract}


\section{Introduction}
A recent 2023 report by the UK Parliament's Women and Equalities Committee on Black maternal health highlights the stark disparities that exist in maternal mortality rates between ethnic groups in the UK \cite{nokes2023black}. Black women are nearly 4 times more likely to die during pregnancy or childbirth than White women. Asian women face almost double the risk of maternal mortality compared to White women. There are also significant differences depending on socioeconomic status, with women in the most deprived areas being 2.5 times more likely to die than those in the least deprived areas. In 2017, the UK government and NHS had set a goal to reduce stillbirths, newborn deaths, maternal deaths, and newborn brain injuries by 50\% by 2025. However, there has been little progress on decreasing maternal mortality rates. When excluding deaths from COVID-19, the maternal mortality rate between 2010-2012 and 2018-2020 increased by 3\% \cite{nokes2023black}. The Healthcare Safety Investigation Branch (HSIB) conducted independent investigations of patient safety incidents in NHS-funded care across England. Established in 2017 and funded by the Department of Health and Social Care, HSIB aimed to improve patient safety through its investigations. HSIB was hosted by NHS England and operated independently. In October 2023, HSIB transformed into two organizations: the Maternity and Newborn Safety Investigations (MNSI), which is hosted by the Care Quality Commission, and the Health Services Safety Investigations Body (HSSIB), which is an independent statutory body. HSIB carried out investigations into adverse incidents during pregnancy and birth. After each investigation, they produced a report discussing the investigation findings and recommendations. These reports were intended to provide insights and lessons learned for mothers and families affected. Maternity investigation work continues at MNSI. 

To enable human factors analysis of maternity incident investigations, we propose the Intelligence Safety Incident Reporting and Annotation (I-SIRch) framework. I-SIRch comprises computational methods and machine learning algorithms to prepare and automatically annotate unstructured text from investigation reports using the Safety Intelligence Research (SIRch) taxonomy, which provides a systematic methodology for extracting safety insights from healthcare investigations. SIRch is utilized to analyze how work system designs can influence patient safety outcomes. Patient harm often stems from problematic systems and processes of care that patients must navigate. By analyzing issues in these complex systems that determine healthcare delivery and identify opportunities to improve system design and reduce patient safety incidents \cite{HSIBthematic}. Such analysis allows safety insights to be extracted based on human factor concepts rather than just clinical factors (e.g. disease, medication, etc).
 
The increasing adoption of concept annotation tools in healthcare and other sectors stems from their ability to extract insights from unstructured text data. Manual annotation requires significant human effort and expertise. This demand has led to growing interest in automated annotation through machine learning and natural language processing. These techniques can accelerate the labeling of domain-specific concepts in text data, enabling more efficient intelligence extraction, search, and analysis. Several concept annotation tools have been specifically designed for the clinical field, including MetaMap \cite{aronson2010overview}, NCBO Annotator \cite{jonquet2009open}, cTAKES \cite{savova2010mayo}, Biomedical Named Entity Recognition (BINER) \cite{Asghari2022}, UniversalNER \cite{zhou2023universalner}, MedCat \cite{kraljevic2021multi}, and BioBERT \cite{lee2020biobert}. These tools leverage medical ontologies to detect clinical concepts within unstructured text and have shown the potential to automate parts of the concept annotation process. These clinical concept annotation tools enable the extraction of biomedical insights but overlook human factors involved in healthcare delivery. Furthermore, significant challenges remain in developing fully automated annotation capabilities. Hybrid human-AI systems provide a promising approach for concept annotation by combining complementary strengths. The accuracy of human experts can be paired with the scalability of AI to efficiently generate high-quality labeled datasets. However, developing robust AI-powered annotation tools introduces several technical challenges. Specifically, large annotated corpora, sufficient computational resources, and ongoing human validation are required to train and evaluate these machine learning models. 

With thoughtful co-design, machine learning automation and human intelligence can build upon each other to enable continued progress in extracting valuable insights from unstructured textual data. We aim to demonstrate the feasibility of automatically annotating incident reports with the SIRch taxonomy, representing a novel contribution as no other model currently exists for this application. Unlike existing clinical annotation models, I-SIRch annotates text based on human factor concepts rather than clinical entities alone. This will provide new insights into how human and system-level issues contribute to patient safety incidents, complementing traditional clinically-focused analyses. By showing the potential of human factors annotation using SIRch, we also open new research directions for understanding and improving socio-technical systems involved in healthcare delivery. In particular, the key contributions of this article are:
\begin{itemize}
\item Provides a computational framework, I-SIRch, that automatically annotates sentences (and text segments) with multiple applicable concepts. The framework was trained on sentences extracted from maternity investigation reports that were manually annotated by patient safety experts. The experts assigned relevant human factor concept(s) from the SIRch taxonomy to each sentence. This mapping enabled the model to learn links between the incidents discussed and socio-technical factors implicated.
\item The I-SIRch framework enables human-AI collaboration, allowing the machine learning model to continuously improve its knowledge by learning from new human expert annotations. This human-in-the-loop process helps develop a robust, adaptable model for extracting insights from reports by leveraging both human expertise and the model's growing knowledgebase.
\item I-SIRch was tested on: 818 synthetically generated sentences; and thereafter an additional 1960 sentences extracted from 97 real reports, to evaluate its generalizability on unseen data. The synthetic sentences were semantically similar to those found in real reports. The real sentences were extracted from healthcare reports that contained sentences that were not previously seen by the model (i.e. not used for training). 
\item Statistical analysis was performed on the annotated text across demographic groups. The annotation results were analyzed to identify insights and differences between ethnic groups regarding factors contributing to incidents. 
\end{itemize}

\section*{Methods}\label{proposedmethosd}
\subsection*{Proposed I-SIRch framework}
\begin{figure}
	\centering
	\includegraphics[width=0.7\textwidth]{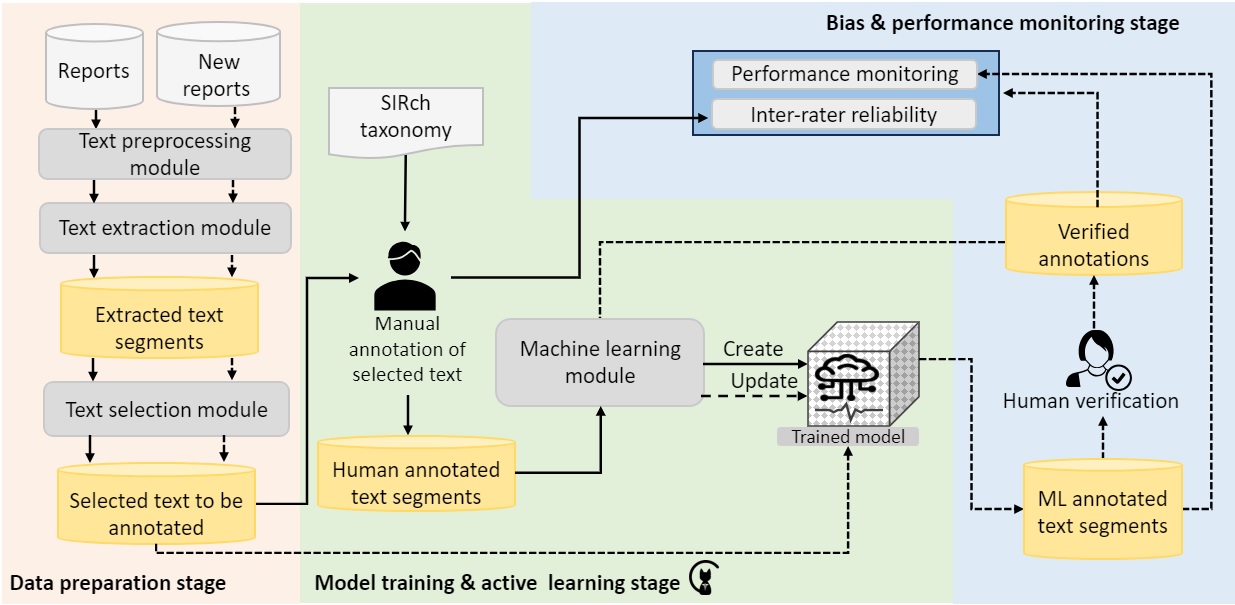}
	\caption{Proposed I-SIRch framework}
 \label{Figure5}
\end{figure}

This section presents the machine learning based framework, I-SIRch, designed for the automatic annotation of maternity investigation reports using the human factors taxonomy, SIRch. Figure \ref{Figure5} illustrates the key components and stages of the I-SIRch framework, depicting its ability to develop a machine learning model capable of learning and adapting to additional inputs over time. A description of each component of the I-SIRch framework is provided below.  

\subsubsection*{Data preparation stage}
\label{sec:dataPreparation}

\paragraph{\textbf{Reports.}} The process starts with a dataset containing maternity investigation reports in PDF format. These reports are introduced into I-SIRch for initial training and subsequently processed through the various modules, as indicated by the solid line in Figure \ref{Figure5}.

\paragraph{\textbf{New reports.}} These reports are different from those used during the initial model training stage. They are introduced into I-SIRch post-training and subsequently processed through the various modules, as indicated by the dashed line in Figure \ref{Figure5}.

\paragraph{\textbf{SIRch taxonomy.}} SIRch codifies and combines the internationally recognized Systems Engineering Initiative for Patient Safety (SEIPS) method \cite{Carayon196}, \cite{Holden2013}, \cite{Carayon2006} with the incident categories used by NHS England and NHS Improvement’s Learn from Patient Safety Events (LFPSE) service \cite{NHSEng2021a}. The SIRch taxonomy that was utilized for training the machine learning model is shown in Supp. File SF1.

\paragraph{\textbf{Text preprocessing module.}} This module's purpose is to preprocess PDF files by eliminating unrecognizable fonts and symbols. It takes a collection of reports located in the `Reports' repository as input. In the process, it extracts and decrypts text from PDF files using Unicode Transformation Format 8-bit (UTF-8) encoding, ensuring readability. It then eliminates unnecessary symbols such as inverted commas (`), dots (.), and unrecognized symbols. Subsequently, the extracted text undergoes cleaning to remove any extraneous sections or elements that do not contribute to the analysis. This accurate extraction of text from PDFs facilitates subsequent natural language processing and makes the content accessible in a standardized machine-readable format. Following the preprocessing, the module stores the processed text in a file for subsequent processing by the `Text extraction module'.

\paragraph{\textbf{Text extraction module.}} The purpose of this module is to extract text from reports based on specific criteria, such as sections, pages, paragraphs, or the entire report. It takes a set of reports as input from the `Text preprocessing module'. The user defines the target section of the report to be extracted. The `text extraction' module then carries out the extraction based on the user's selection, whether it is a particular section, page, paragraph, or entire report. It is important to note that in structured reports, the user specifies a criterion for text selection (e.g., a section within an investigation report), which will be extracted from all reports, while in unstructured documents, the user has the option to extract text from the entire document. Following the extraction process, the module stores the extracted text in a file, preparing it for further processing by the `Text selection module'.

\begin{figure}
	\centering
	\includegraphics[width=0.65\textwidth]{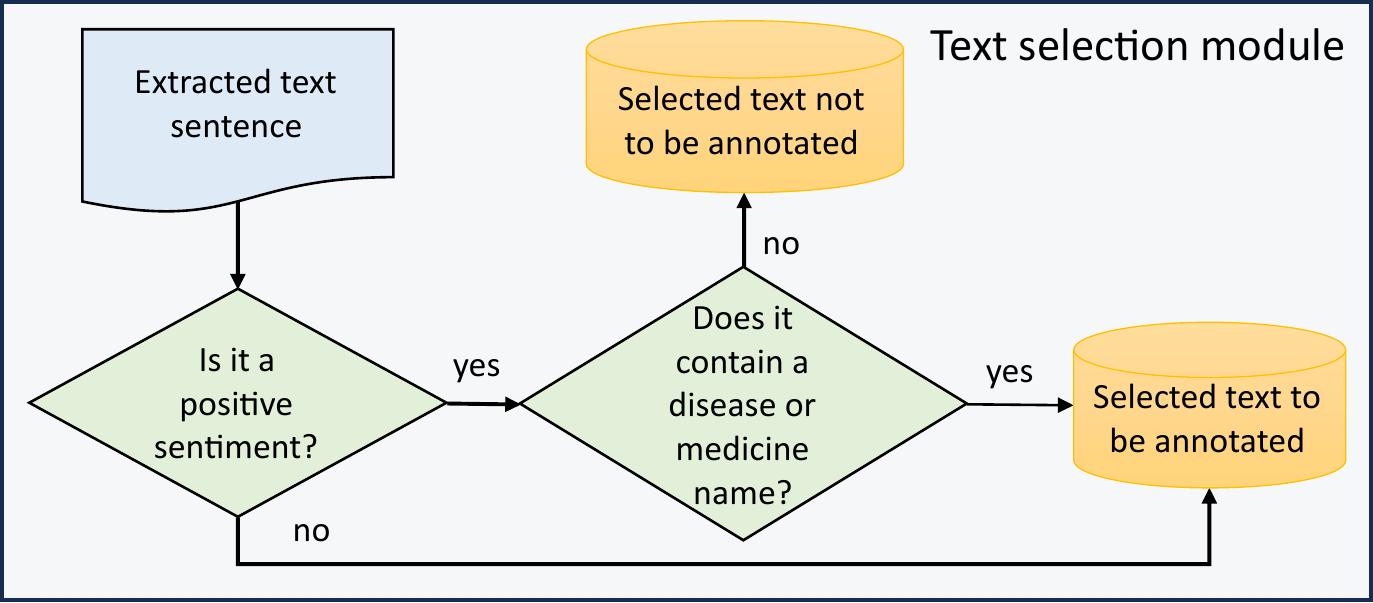}
	\caption{Text selection module}
 \label{Figure6}
\end{figure}

\paragraph{\textbf{Text selection module.}} The purpose of this module is to identify and select sentences with negative connotations, references to physical characteristics, and medicine names (dispensing medication). It takes the extracted text as input from the `Text extraction module'. Initially, it uses a deep learning transformer model that was fine-tuned for identifying negated sentences. The module also maintains a list of negation keywords, including the terms `not' and `never', to catch any negated cases that might be missed by the model. Additionally, the module scans for the phrase `in line with' in positive sentences and marks tokens following it as affirmed. Any sentence found to possess a negative meaning or contain references to physical characteristics or medication names is flagged for annotation. Subsequently, the module stores the selected sentences in a file entitled `Selected text to be annotated', preparing them for further training (i.e. retraining) by the `Machine learning module'. Figure \ref{Figure6} depicts the process of how sentences are selected for automatic concept annotation.  

\paragraph{\textbf{Manual annotation of selected text.}} In this process, multiple human experts independently annotate the same set of selected sentences using the SIRch framework. This approach enables the evaluation of inter-rater reliability (IRR) and the consistency of annotations across various experts. These experts manually annotate (or code) the selected segments that highlight essential investigation findings. They accomplish this by assigning pertinent concepts from the specified taxonomy to the sentences or individual words. The annotation process can make use of the MedCATtrainer \cite{kraljevic2021multi} tool. An illustrative example of manual annotation is provided in Supp. Table S12. The resulting manually annotated sentences are systematically stored in a repository named `Human annotated text segments'. These annotations serve as training data for the `Machine learning module'.

\subsubsection*{Model training and active learning stage}
\paragraph{\textbf{Machine learning module.}}  This module is responsible for training a natural language processing (NLP) model with the ability to automatically annotate concepts within selected text segments. It takes as input sentences manually annotated by human experts, stored in the `Human annotated segments' repository. The `machine learning module' encompasses several steps. Initially, it establishes a concept database (CDB) containing the SIRch concepts to be recognized. The `Human annotated text segments' are then uploaded into the MedCATtrainer \cite{kraljevic2021multi}, where human experts use its graphical user interface to annotate the text segments. Once MedCATtrainer learns from a few annotated examples, it can make predictions, facilitating faster annotation through active learning. A concept needs to appear at least once in the training data to be considered for training. Human experts utilize MedCATtrainer \cite{kraljevic2021multi} to validate and correct automatic annotations. These validated examples contribute to fine-tuning the model, promoting an active learning process. The module's output is a `Trained model' that is stored and ready for deployment in automatic annotation tasks, simplifying the recognition of concepts within textual data. When the trained model is employed to make predictions on previously unseen reports, the annotated sentences are stored in the `ML annotated text segments repository and subsequently utilised in the `Bias and performance monitoring' stage. 

\subsubsection*{Bias and performance monitoring stage} 
\paragraph{Human verification.} The purpose of this stage is to assess the correctness of annotations generated by the trained model. It begins by taking the sentences annotated by the trained model as input for human evaluation. The process involves several steps. Initially, a new dataset is passed through the I-SIRch pipeline and processed through the `Trained model,' which annotates the selected text. These annotated sentences are stored in the `ML annotated text segments' repository. Human experts then access this repository to manually verify whether the model's predicted annotations for each sentence are correct or incorrect. These verified labels, indicating correctness or errors, are recorded as `Verified annotations'. Any sentences with errors are isolated and used to retrain the model, thereby enhancing its performance. The updated model, after retraining with the verification data, becomes ready for deployment on new sentences. This verification-retraining loop allows for continuous performance monitoring and improvement of the model. Finally, the verified annotations (i.e. concepts) are passed on to the `Bias and performance monitoring stage', where a more comprehensive evaluation of the model's biases, errors, and real-world effectiveness is conducted to ensure rigorous monitoring and ongoing enhancement of the model's performance and reliability.

\paragraph{\textbf{Performance monitoring.}} The purpose of this module is to evaluate the performance of the trained machine learning model. This evaluation takes as input a dataset comprising verified human annotations, which forms the foundation for rigorous analysis. The process entails the calculation of various performance metrics, including balanced accuracy, precision, recall, and F-score, providing a comprehensive assessment of the model's concept annotation performance. Moreover, the model undergoes scrutiny for biased performance across diverse demographic groups or sentence types, ensuring fairness and impartiality. For instance, its performance is examined for systematic variations when annotating sentences from different ethnic groups.

\paragraph{\textbf{Inter-rater reliability (IRR).}} Additionally, inter-rater reliability (IRR) is calculated by comparing annotations from multiple human experts for the same sentences. High IRR indicates consistent human annotation, whereas low IRR may indicate ambiguity within the taxonomy. Ongoing monitoring of both model performance metrics and IRR provides valuable insights into potential refinements in the taxonomy, annotation process, and model training approach. This iterative process, continuously adapting to new data, allows for ongoing improvements and updates to the model, while effectively addressing biases. The end goal is the development of a machine learning model that leverages both expert-annotated data and its own predictions to continually learn and enhance its text classification capabilities.

\subsection*{Experiment Methodology}
Four data batches were utilised for training and testing I-SIRch. Supp. Table S1 holds information about the batches, and Supp. Table S13 shows the frequencies of concepts found in the batches that were utilized for developing the model.

\begin{itemize}[leftmargin=1in]
\item [Training:] The maternity incident reports (Batch 1) provided by HSIB were utilised for training the model, following the process described in Section \textit{I-SIRch framework}. With regards to the `Human annotation' stage, there were two experts from HSIB and a human factors expert from Loughborough University annotating the reports. Of those 76 maternity investigation reports, 20 reports (sentences $= 184$) were randomly selected for annotation by three human experts as discussed in the inter-rater reliability (IRR) step of the \textit{Bias and performance monitoring stage}. The observed IRR among three annotators was 80.15\% (see Supp. Table S14). Ethnic groups were unknown and not considered during the human annotation stage to reduce any potential biases entering the system during human annotation. Performance was evaluated by comparing the predicted vs actual (i.e. relevant) concepts.
\item [Testing (Test A):] The synthetic (test) dataset (Batch 4) that was generated following the process described in Supp. File SF2, served as the set for testing the performance of the trained model on a previously unseen but semantically similar dataset. The relevant concepts for each sentence were known since they were mapped to the concepts of those sentences extracted from the real dataset. Ethnicity was considered when analysing the performance of the model across the various ethnic groups.
\item[Retraining 1:] After testing with the synthetic dataset (Batch 4), a total of 15 real reports were utilized to further train the model (Batch 2). These 15 reports were selected because they contained the less frequent codes that were identified from the previous dataset. The I-SIRch model was utilized to extract and automatically annotate 344 sentences from these reports, and thereafter, a human annotator verified the correctness of the annotations. 
\item [Retesting 1 (Test B):] Thereafter a total of 97 real reports (Batch 3) were utilized for further testing. The I-SIRch model was utilised to automatically annotate 1960 sentences extracted from the reports, and thereafter a human annotator verified the correctness of the annotations. This enabled for evaluation of performance with regards to applying the trained model on a real unseen dataset. Ethnicity and healthcare outcomes were not provided with these reports. Retesting 1 is referred to as Test B.
\item [Retraining 2:] The 97 reports (Batch 3) were then used for retraining the model. 
\item [Retesting 2 (Test C):] After retraining the model with Batch 3, the model was tested using the synthetic dataset (Batch 4).
\end{itemize}

\subsubsection*{Metrics for evaluating the performance of machine learning based annotation}

\label{sec:evaluation}
A set of evaluation metrics was employed to assess the trained model's performance in automatically annotating sentences from the test sets. In the context of a concept annotation task, the evaluation of system or annotator performance involves considering four key metrics: True Positives (TP), False Positives (FP), False Negatives (FN); and True Negatives (TN). These are described below, and provide the basis for calculating Recall, Precision, and the $F$-Score to assess the performance of the system.

\paragraph{\textbf{True Positive (TP).}} is the count of concepts in a sentence that were correctly identified and annotated as a specific concept. In other words, these are instances where the system correctly recognised and marked the concept where it truly exists in the sentence.

\paragraph{\textbf{False Positive (FP).}} is the count of concepts that were incorrectly annotated in a sentence, falsely identified as part of the concept when they are not. These are instances where the system made an error by including concepts that should not have been part of the annotated concept.

\paragraph{\textbf{False Negative (FN).}} denotes the count of concepts that should have been annotated as part of the concept in a sentence but were missed or omitted during the annotation process. These are instances where the system failed to recognize and include concepts that should have been part of the concept.

\paragraph{\textbf{True Negative (TN).}} is the count of concepts that were not annotated because they should not have been included in the concept. These are instances where the system correctly identified that certain concepts were not relevant to the annotated concept and left them out.

\paragraph{\textbf{Recall.}} measures the proportion of actual annotations that are correctly identified by the system. It assesses how comprehensive the system is in capturing all the relevant annotations. A higher recall indicates that the system can identify a greater number of actual annotations, although it may also include some false positives.
  \begin{equation}
        Recall = TP/(TP+FN), \in(0,1)
    \end{equation}
High recall indicates that the system or annotator is good at capturing all instances of the concept in the text.
\paragraph{\textbf{Precision.}} measures the proportion of predicted annotations that are actually correct. It indicates how accurately the system can identify the relevant annotations without mislabeling irrelevant ones. A higher precision indicates that the system is more reliable in identifying the correct annotations.
    \begin{equation}
        Precision = TP/(TP+FP), \in(0,1),
    \end{equation}
High precision indicates that when the system or annotator claims a code is part of the concept, it is highly likely to be correct.

\paragraph{\textbf{F-score.}} is a balanced measure that takes into account both precision and recall in a concept annotation task. It is the harmonic mean of precision and recall, providing a single metric that balances the trade-off between ensuring all relevant concepts are included in the annotated concept (high recall) while maintaining high precision:
    \begin{equation}
        F-score = 2\times Precision \times Recall/(Precision + Recall), \in(0,1)
    \end{equation}    
The $F$-Score is particularly useful when aiming to strike a balance between ensuring that all relevant concepts are included in the annotated concept (high recall) while maintaining high precision. It serves as an overall measure of the effectiveness of the concept annotation system or annotator.

\paragraph{\textbf{Accuracy.}} measures the overall correctness of the annotations made by a system. It is defined as the ratio of correctly annotated concepts (True Positives and True Negatives) to the total number of annotations (True Positives + True Negatives + False Positives + False Negatives), as follows.
    \begin{equation}
        Accuracy = (TP+TN)/(TP+FP+TN+FN), \in(0,1)
    \end{equation} 
Accuracy provides an overall assessment of how well the system or annotator is performing in correctly identifying both positive and negative cases within the annotated concepts, thus quantifying the system's annotation correctness.

\paragraph{\textbf{Balanced Accuracy.}} Accuracy can be a misleading metric for imbalanced data sets. Given that some concepts are more frequent than others in the dataset, it is important to also report the balanced accuracy as a measure of accuracy. The balanced accuracy is the average between the sensitivity and the specificity, which measures the average accuracy obtained from both the minority and majority classes. The balanced accuracy measure was calculated as follows. First calculating the True Positive Rate (TPR) and True Negative Rate (TNR) as 
$TPR=TP/(TP+FN)$ and $TNR= TN/(TN+FP)$, respectively. Balanced Accuracy is thereafter calculated as follows.
 \begin{equation}
Balanced Accuracy=(TPR+TNR)/2, \in(0,1)
\end{equation}

\paragraph{\textbf{Inter-rated reliability (IRR)}} is the degree of agreement among the human experts who annotated the initial reports before being used for initial model training. IRR results are shown in the Supp. Table S14. IRR is calculated as follows.
\begin{equation}
    IRR = \frac{Total\ number\ of\ concept \ agreements}{Total\ number\ of\ concepts\ \times Number\ of\ annotators}
    \label{eqn:IRR}
\end{equation}

\section*{Results}
\subsection*{Dataset of maternity incident investigation reports} 

HSIB provided a random set of 188 investigation reports describing adverse maternity incidents (see Batches 1-3 in Supp. Table S1). The reports were written between 2019 to 2022. The number of reports for each year is as follows: 4 reports in 2019, 115 reports in 2020, 42 reports in 2021, and 27 reports in 2022. Ethnicity was only provided for Batch 1 which comprised 76 reports. The discussion that follows focuses on Batch 1, since ethnicity was not available for the other batches. 

\begin{figure}
	\centering
\includegraphics[width=0.6\textwidth]{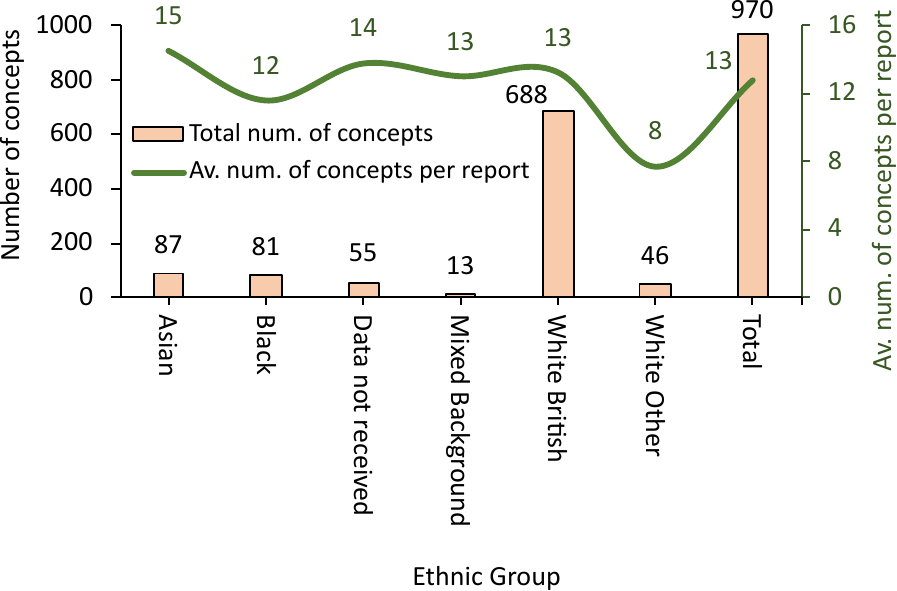}
	\caption{Total number of concepts and average number of concepts per report across ethnic groups.}
 \label{Figure1}
\end{figure}

\begin{figure}
	\centering
\includegraphics[width=0.6\textwidth]{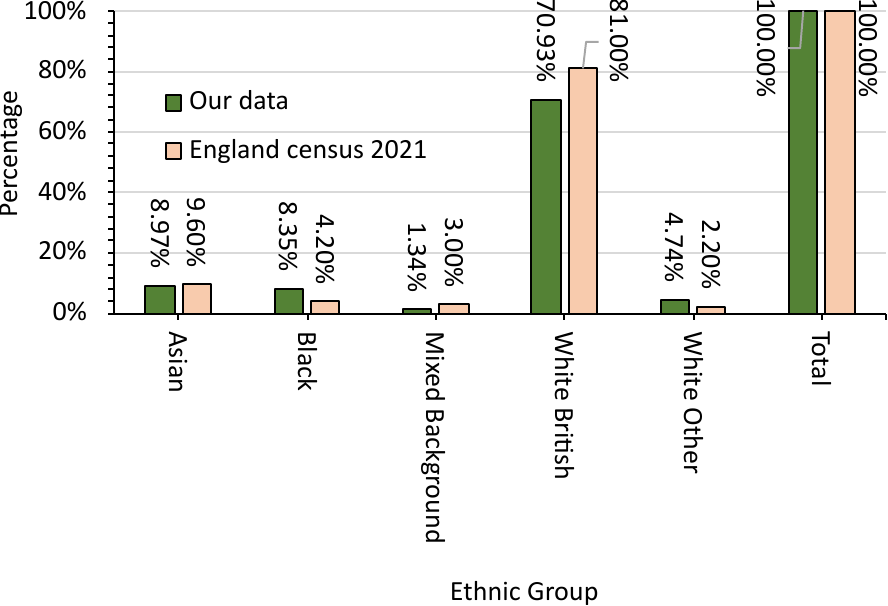}
	\caption{Percentage of mothers and families from each ethnic group in our dataset compared to England's 2021 census data.}
 	\label{Figure2}
\end{figure}

\paragraph{Batch 1 analysis.} The percentage of reports containing each concept/human factor across each ethnic group is shown in Supp. Table S2. Figure \ref{Figure1} illustrates a description of the number of reports and concepts (human annotations using SIRch) found within the reports of Batch 1. The corresponding data can be found in Supp. Table S3.
As shown in Figure \ref{Figure2}, 70.93\% of the reports are of the White British ethnic group, compared to 81\% nationally. The reports also cover 8.97\% Asian (vs. 9.6\% nationally) and 8.35\% Black (vs. 4.2\% nationally) ethnic groups. The distribution of ethnic groups in the dataset closely mirrors the population of England based on the 2021 census \cite{Garlick_2022}. This indicates that the reports provide a fairly representative sample across various ethnic groups in England, with a reasonable representation of the Black ethnic group compared to the national population\cite{Garlick_2022} (Figure \ref{Figure2}). The frequency distribution of concepts across reports for each ethnic group are provided in Supp. Table S4. 

Supp. Table S5 provides a summary of the 76 reports describing adverse maternal and neonatal outcomes, categorized by ethnicity. These outcomes include instances of babies who received therapeutic hypothermia, early neonatal deaths, maternal deaths, and intrapartum stillbirths. Within the dataset, the most common adverse incident for reported cases for mothers of the White British ethnic group was babies receiving therapeutic hypothermia (30 out of 52 reports of their reports), whereas, for mothers of the Black ethnic group, it was maternal death (4 out of 7 of their reports). Please note that the dataset is a random set of reports extracted for testing the proposed concept annotation approach and demonstrating how the results can be analyzed, and it is not a representation of the reports held by HSIB.

\subsection*{Model performance evaluation on the synthetic (test) set across ethnic groups}

Supp. Table S1 shows the batches utilized for retraining and testing the model. The methodology section describes the retraining and testing process that was followed. I-SIRch was tested three times as follows: \textbf{Test A:} Training on Batch 1 (real-data) and testing on Batch 4 (synthetic data. \textbf{Test B:} Retraining on Batch 2 (real data), and testing on Batch 3 (real data). \textbf{Test C:} Retraining again on Batch 3 (real data) and testing on Batch 4 (synthetic data). The results of all tests are shown in Supp. Tables S6 and S7. 

\paragraph{\textbf{Test A.} } The I-SIRch model was initially trained on Batch 1 and tested on Batch 4. Table \ref{Table1} shows the performance of the model when using the metrics described in the Experiment Methodology section. Supp. Table S8 shows the performance of the model across concepts on the synthetic test dataset (Batch 4). Supp. Table S9 (Test A) shows the results of the correct and incorrect annotations per ethnic group.
I-SIRch was initially utilized for automatically coding a test set of 970 synthetic sentences (Batch 4). The process describing how the synthetic sentences were generated is found in Supp. File S12. Metrics were applied to evaluate the performance of the model per sentence and overall model performance. The trained machine learning model demonstrated strong performance on the synthetic test set across all evaluation metrics. 

The precision of 0.87 indicates that the model has a low false positive rate in its annotations. The recall of 0.93 indicates effective identification of appropriate concepts with minimal false negatives. Finally, the F-score of 0.96 validates the overall effectiveness of the model by balancing both precision and recall. The high F-score indicates that the model was able to apply concepts where relevant while avoiding incorrect annotations. In summary, the high precision, recall, and F-score reflect the model's capability at multi-label annotation of sentences using the SIRch taxonomy. Further tests to explore the feasibility of the I-SIRch concept annotation framework on real data are described later on in the paper (Tests B and C).
\newpage
\begin{table}
\begin{longtable}[c]{lccccc}
\caption{Performance of I-SIRch when trained on Batch 1 and tested on Batch 4 (Test A). Table shows the mean and standard deviation values.
    } \label{Table1}\tabularnewline
        \hline 
        Ethnic group & Precision & Recall & F-score & Misc. & Bal. Acc.\tabularnewline
                \hline 
        Asian & 1.00 \textpm{} 0.00 & 0.60 \textpm{} 0.23 & 0.79 \textpm{} 0.09 & 0.40 \textpm{} 0.23 & 0.80 \textpm{} 0.11\tabularnewline
        Black & 1.00\textpm{} 0.00 & 0.59 \textpm{} 0.21 & 0.78 \textpm{} 0.08 & 0.41 \textpm{} 0.21 & 0.80 \textpm{} 0.10\tabularnewline
        Data not received & 1.00 \textpm{} 0.00 & 0.59 \textpm{} 0.24 & 0.80 \textpm{} 0.08 & 0.41 \textpm{} 0.24 & 0.80 \textpm{} 0.12\tabularnewline
        Mixed Background & 1.00\textpm{} 0.00 & 0.65 \textpm{} 0.12 & 0.78 \textpm{} 0.09 & 0.35 \textpm{} 0.12 & 0.82 \textpm{} 0.06\tabularnewline
        Other White & 1.00 \textpm{} 0.00 & 0.58 \textpm{} 0.25 & 0.79 \textpm{} 0.09 & 0.42 \textpm{} 0.25 & 0.79 \textpm{} 0.12\tabularnewline
        White British & 1.00 \textpm{} 0.00 & 0.60 \textpm{} 0.23 & 0.79 \textpm{} 0.08 & 0.40 \textpm{} 0.23 & 0.80 \textpm{} 0.11\tabularnewline
        \hline
        Average & 1.00 \textpm{} 0.00 & 0.60 \textpm{} 0.21 & 0.78 \textpm{} 0.08 & 0.40 \textpm{} 0.21 & 0.80 \textpm{} 0.10\tabularnewline
        \hline 
        \multicolumn{6}{l}{Bal. Acc.: Balanced Accuracy; Misc: Misclassification}\tabularnewline
        \hline 
\end{longtable}
\end{table}

\begin{figure}
	\centering
\includegraphics[width=0.60\textwidth]{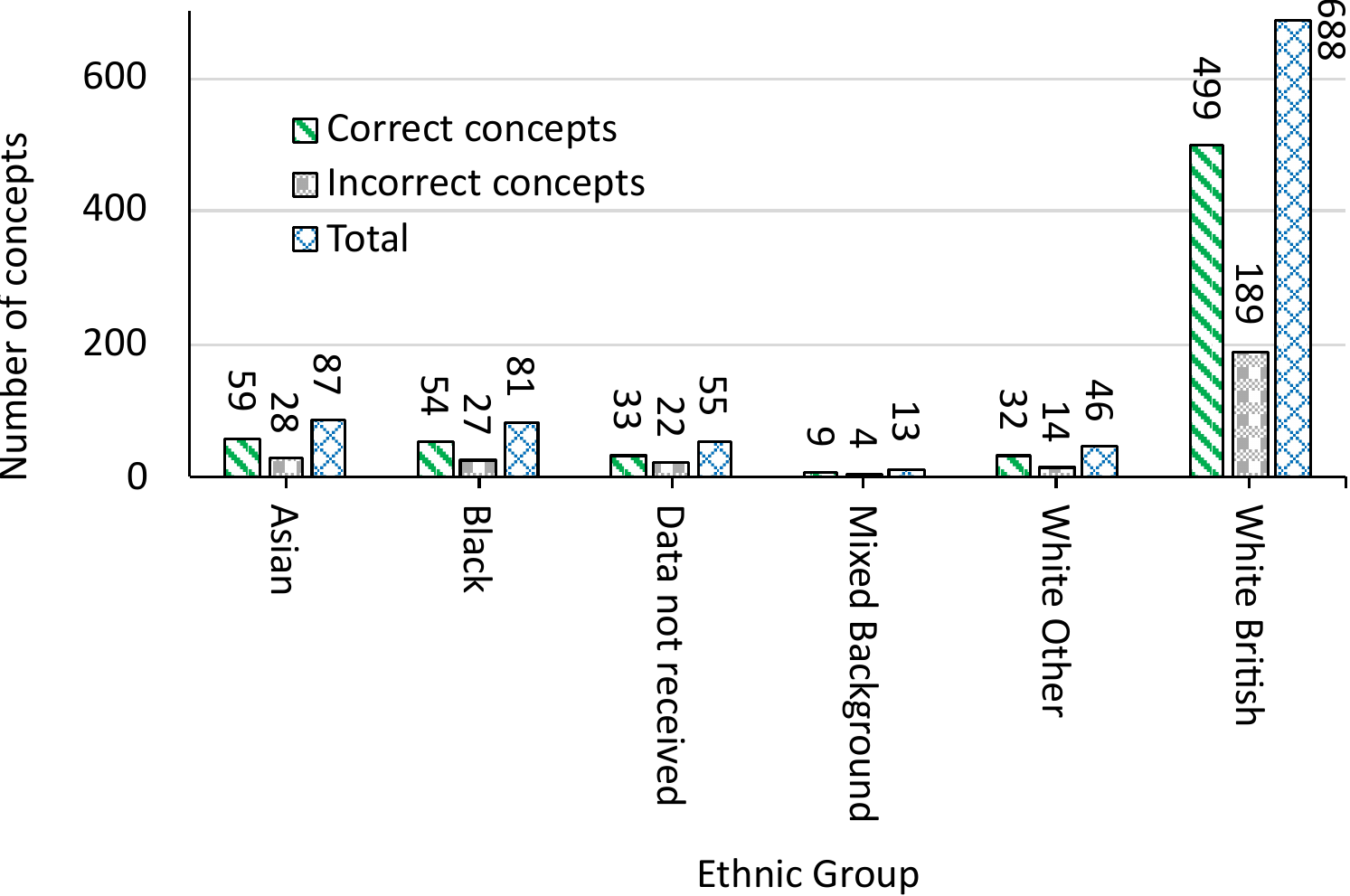}
	\caption{Average number of correct and incorrect concepts across various ethnic groups (Batch 1).}
\label{Figure3}
\end{figure}

\paragraph{\textbf{Performance across ethnic groups (Test A).}} The synthetic dataset contains 818 sentences (970 concept annotations), with the distribution across ethnic groups matching those of Batch 1 (Figure \ref{Figure1}). The distribution of ethnic groups of the synthetic dataset maps to the real datasets because the synthetic sentences were generated from real sentences extracted from reports for which the ethnicity was known. The model correctly annotated 70.93\% of sentences, with 686 correct annotations out of a total of 970 (see Supp. Table S9). The model performed well across groups when ethnicity was considered (see Figure \ref{Figure3}), showing no significant accuracy (here accuracy refers to the percentage of correctness in identifying the TP annotations) discrepancies, as discussed in the following subsection. As shown in Supp. Table S9, the sentences of the reports of the White British ethnic group saw the highest accuracy of 72.82\%, with 499 correct annotations out of 688. For minority ethnic groups, the model obtained accuracy values of 67.82\% on Asian examples (59 correct out of 87), 66.67\% on Black examples (54 correct out of 81), and 69.23\% on the smaller Mixed Background set (9 out of 13). For the `White Other' ethnic group, the model reached 69.57\% accuracy (32 out of 46). Lower performance (i.e. 60\% accuracy, with 33 correct out of 55) was observed in annotations where `ethnicity data was not received', highlighting an area for improvement. However, the comparable accuracy across ethnic groups generally demonstrates the model's capability to learn meaningful patterns from diverse ethnic groups without significant biases. Analysing by ethnic group provides valuable insights into model fairness, and the results indicate strong generalizability.

\subsection*{Are there any significant differences in the performance of I-SIRch on test data, across the Black and White ethnic groups?}

A Wilcoxon signed-rank test was conducted to determine whether there were significant differences in the performance of I-SIRch's machine learning model between the Black and White British ethnic groups. The results showed no significant difference in model performance between Black (median = 66.67\%, SD = 30.37\%) and White British (median = 70.14\%, SD = 7.5\%) ethnic groups, $Z =-0.806$, $p = 0.42$. Based on these findings, it can be concluded that the model demonstrated no significant variation in performance across the Black and White British ethnic groups analyzed. The Wilcoxon signed-rank test indicates that the model performance was statistically comparable for the Black and White British ethnic groups represented in the dataset. However, the higher standard deviation for the Black ethnic group points to greater variability in model performance compared to the White British group. This suggests a more limited or inconsistent representation of text from the reports of the Black ethnic group. Expanding the diversity of the training data can enhance the model's ability to make fair and accurate predictions without unintended biases related to ethnicity, and can reduce performance fluctuations across groups. Ongoing evaluation of the model will be performed and retrained over time to check for potential biases which is key to maintaining fairness as the model evolves. 

\subsection*{Can performance improve when retraining the model with real data, and testing on real and synthetic data?} 

\paragraph{Test B.} After the Retraining 1 phase (using Batch 2), the performance of the model was evaluated on a real dataset (Batch 3) comprising 1960 sentences extracted from 97 reports. The results are shown in Supp. Table S6 (Test B). Note that ethnicity was not available for Batch 3 and hence a performance analysis of the model across the ethnic groups could not be conducted. The I-SIRch model reached the highest balanced accuracy of 0.90 $\pm 0.18$ when tested on real data. The performance of I-SIRch on Test B was an improvement compared to its performance when tested on the synthetic dataset during tests A and B, with balance accuracy values of 0.80 $\pm 0.11$ and 0.83 $\pm 0.08$, respectively.

\begin{figure}[!h]
\begin{subfigure}[b]{.5\textwidth}
\centering
\includegraphics[width=1\textwidth]{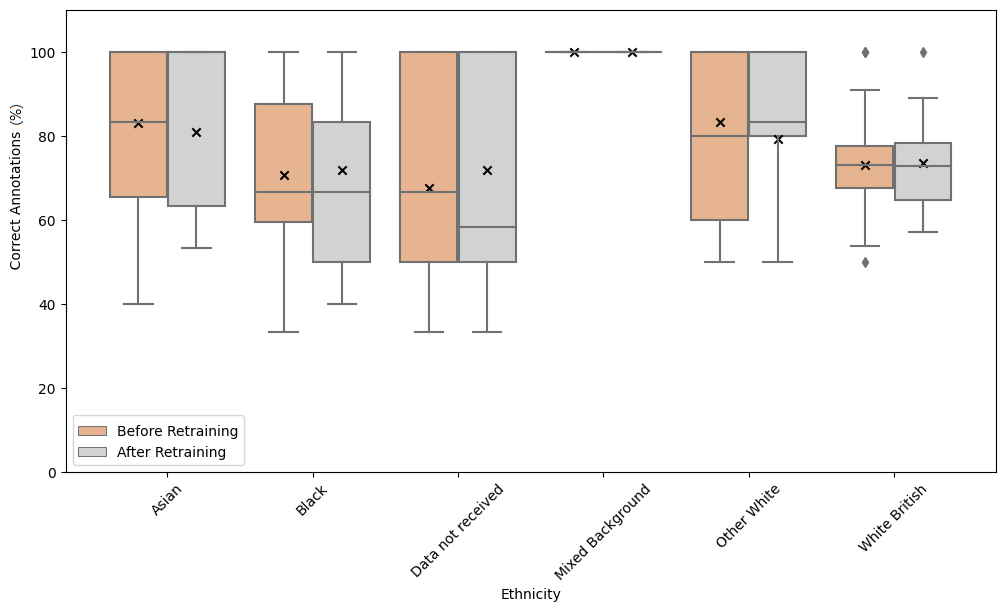}
\caption{Distribution of correct annotations for the various ethnic groups}
\label{Figure4a}
\end{subfigure}
\hfill
\begin{subfigure}[b]{.5\textwidth}
\centering
\includegraphics[width=1\textwidth]{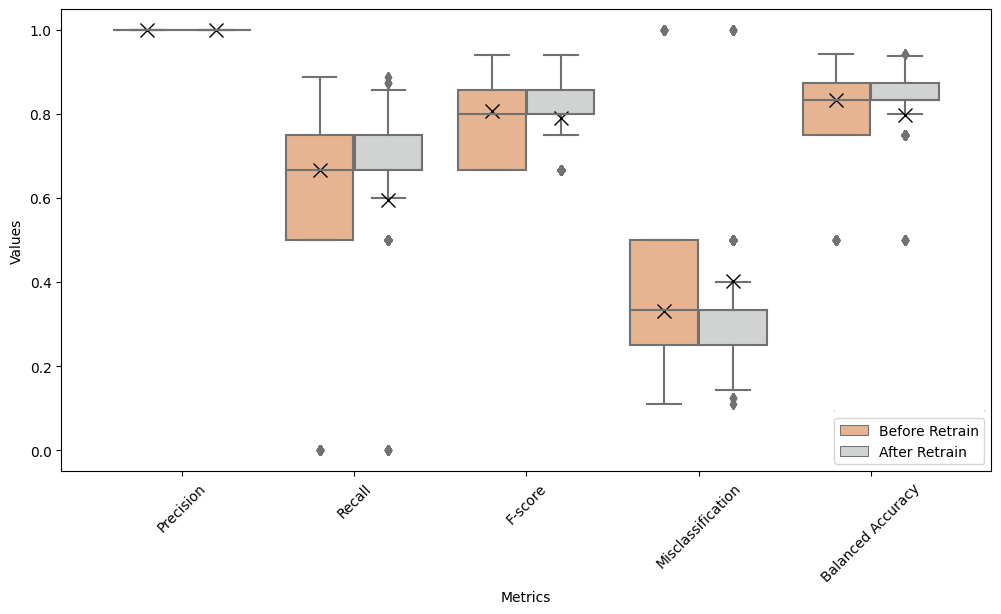}
\caption{Performance evaluation using various metrics }
\label{Figure4b}
\end{subfigure}
\caption{\textbf{Orange bars} show the test results when the model is trained on Batch 1 and tested on Batch 4 (Test A). \textbf{Grey bars} show the test results when the model is trained on Batch 1, retrained using Batches 2 and 3 and tested on Batch 4 (Test C).}
\label{Figure4}
\end{figure}

\paragraph{Test C.} This test aims to evaluate the performance of the I-SIRch model when trained using all the available real data (Batches 1, 2, and 3) and tested on the synthetic data (Batch 4). Hence, during Retraining phase 2, Batch 3 was utilized to further retrain the model (i.e. model has already been trained with Batches 1 and 2), and testing was conducted using the synthetic data (Batch 4). Supp. Table S10 shows the performance of I-SIRch when tested across each concept and ethnic group. Table \ref{Table2} shows the average results across the ethnic groups when using various evaluation metrics. The results of Test A and Test C can be directly compared because both tests were conducted on the synthetic dataset (Batch 4). Test B was conducted on Batch 3 and hence cannot be directly compared to the results of Test A and Test C. Comparing the results of Tests A and C in Table S6 and Figure \ref{Figure4}, there is an improvement in the performance of the model after retraining the model using Batch 2. The results of a Wilcoxon signed-rank test (see  Supp. Table S11) show that this improvement is not statistically significant. It is however worth noting that Table S9 shows that there was an increase in the number of correctly annotated concepts across all ethnic groups.

\newpage
\begin{table}[!h]
\begin{longtable}[c]{lccccc}
\caption{Performance of I-SIRch when tested on the synthetic data (batch 4) after training on Batch 1 and retaining on Batches 2 and 3. Performance across the ethnic groups. Table shows the mean and standard deviation values. (Test C)} \label{Table2}\tabularnewline
    \hline 
    Ethnic group & Precision & Recall & F-score & Misc. & Bal. Acc.\tabularnewline
    \hline
    Asian & 1.00 \textpm{} 0.00 & 0.68 \textpm{} 0.14 & 0.81 \textpm{} 0.09 & 0.32 \textpm{} 0.14 & 0.84 \textpm{} 0.07\tabularnewline
    Black & 1.00 \textpm{} 0.00 & 0.65 \textpm{} 0.17 & 0.80 \textpm{} 0.07 & 0.35 \textpm{} 0.17 & 0.82 \textpm{} 0.09\tabularnewline
    Data not received & 1.00 \textpm{} 0.00 & 0.65 \textpm{} 0.21 & 0.82 \textpm{} 0.08 & 0.35 \textpm{} 0.21 & 0.82 \textpm{} 0.10\tabularnewline
    Mixed Background & 1.00 \textpm{} 0.00 & 0.69 \textpm{} 0.12 & 0.81 \textpm{} 0.09 & 0.31 \textpm{} 0.12 & 0.85 \textpm{} 0.06\tabularnewline
    Other White & 1.00 \textpm{} 0.00 & 0.64 \textpm{} 0.19 & 0.80 \textpm{} 0.09 & 0.36 \textpm{} 0.19 & 0.82 \textpm{} 0.09\tabularnewline
    White British & 1.00 \textpm{} 0.00 & 0.67 \textpm{} 0.15 & 0.81 \textpm{} 0.08 & 0.33 \textpm{} 0.15 & 0.84 \textpm{} 0.07\tabularnewline 
    \hline
    Average & 1.00 \textpm{} 0.00 & 0.66 \textpm{} 0.16 & 0.80 \textpm{} 0.08 & 0.34 \textpm{} 0.16 & 0.83 \textpm{} 0.08\tabularnewline
    \hline 
    \multicolumn{6}{l}{Bal. Acc.: Balanced Accuracy; Misc: Misclassification}\tabularnewline
    \hline 
\end{longtable}
\end{table}

\section*{Conclusion and discussion}
This paper introduces I-SIRch, a novel framework for automating the annotation of healthcare investigation reports using human factors concepts. I-SIRch utilizes the Safety Intelligence Research (SIRch) taxonomy developed by England's Healthcare Safety Investigation Branch to systematically analyze how work system factors contribute to patient safety incidents. The key innovation of I-SIRch is enabling the annotation of reports based on human and system-level issues, rather than just biomedical concepts. This provides new insights into the socio-technical dimensions of safety lapses in healthcare delivery. The framework comprises computational methods for extracting, preparing, and annotating unstructured text from investigation reports. A machine learning model was trained to annotate sentences with applicable SIRch concepts that implicate various human factors.

The I-SIRch model was initially trained on 76 real maternity investigation reports manually annotated by experts using the SIRch human factors taxonomy. Inter-rater reliability (IRR) analysis found 80.15\% agreement between three expert human annotators. The model was tested on 818 synthetic sentences automatically generated to be semantically similar to real data. On this set, I-SIRch achieved a balanced accuracy of approximately 80\%. The model underwent two further retraining iterations and was finally tested on 1960 real sentences extracted from 97 previously unseen real reports, attaining a balanced accuracy of approximately 90\% in assigning applicable human factor concepts. Analysis of predicted annotations revealed communication and care coordination issues (i.e. organisation-teamworking) prevalent across ethnic groups, while gaps in risk assessment were greater for minority mothers. Ongoing analysis of model predictions by human experts ensures rigorous performance monitoring and bias detection.

Overall, this work establishes a foundation for extracting intelligence on work system contributors to patient harm, complementing traditional clinical perspectives. As healthcare increasingly embraces AI automation, purposeful integration of human factors alongside biomedical understanding will provide a more comprehensive view of safety vulnerabilities. The I-SIRch concept annotation tool offers an important step toward rigorous, scalable analysis of socio-technical dimensions in health services delivery. The proposed human-in-the-loop framework promotes continuous evolution of the AI model, while meaningfully incorporating human expertise. By extracting detailed insights into how human and system-level factors contribute to incidents, I-SIRch can better inform targeted improvements to complex healthcare systems that impact patient outcomes.

\section*{Acknowledgements}
The work was jointly funded by The Health Foundation and the NHS AI Lab at the NHS Transformation Directorate, and supported by the National Institute for Health Research. The project is entitled ``I-SIRch - Using Artificial Intelligence to Improve the Investigation of Factors
Contributing to Adverse Maternity Incidents involving Black Mothers and Families'' AI\_HI200006. The authors would like to acknowledge MNSI for their feedback on the paper. 

\section*{Author contributions statement}
G.C and M.K designed and conceived the experiments.  M.K and G.C. conducted the experiments. G.C and M.K analyzed the results and wrote the manuscript. P.W and T.J led and performed the human annotation of reports. J.B. provided expertise on safety investigations and the SIRch coding taxonomy. All authors reviewed the manuscript. 

\section*{Data availability}
Anonymised raw data were provided by HSIB. The Supplementary File provides the data that was generated from the incident investigation reports. 

\section*{Conflict of interest}
On behalf of all authors, the corresponding author states that there is no conflict of interest

 \clearpage 




 \captionsetup[table]{name=Supplementary Table S, labelformat=nospace}
 \setcounter{table}{0}  



\captionsetup[table]{name=Supplementary Table S, labelformat=nospace}




\section*{Supplementary Information}
\vspace{-0.4cm}
\begin{table}[!h]
\centering
\caption{Number of concepts and sentences per batch.} 
\label{SupplementaryTable 1}
\begin{tabular}{lcc}
\hline 
No. of reports & no. of concepts & no. of sentences \tabularnewline 
\hline
Batch 1 (n=76 real) & 970 & 818 \tabularnewline
Batch 2 (n=15 real) & 452 & 344 \tabularnewline 
Batch 3 (n=97 real) & 2644 & 1960 \tabularnewline
Batch 4 (n=76 synthetic) & 970 & 818 \tabularnewline 
\hline 
\multicolumn{3}{l}{Reports per year (excluding batch 4): 2019 (n=4), 2020(n=115), 2021 (n=42), 2022 (n=27)}  \tabularnewline
\hline  
\end{tabular}
\end{table}
\vspace{-0.4cm}
 \begin{longtable}{p{7cm}cccccc|cc}
    \caption{Percentage of reports from Batch 1 containing each concept/human factor for each ethnic group calculated by using the formula shown at the bottom of this table. For example, the Organisation-Communication factor appeared in 7 out 7 reports of the Black ethnic group, hence 100\% occurrence.}
    \label{SupplementaryTable 2}\tabularnewline
        \hline 
       Concept  & Asian & Black & DNR & MB & WO & WB & Average & SD\tabularnewline
        \hline 
        Organisation-Communication factor & 83.33 & 100.00 & 75.00 & 100.00 & 66.67 & 75.00 & 83.33 & 13.94\tabularnewline
        External Environment-COVID & 66.67 & 57.14 & 25.00 & 100.00 & 66.67 & 80.77 & 66.04 & 25.04\tabularnewline
        Organisation-Teamworking & 83.33 & 71.43 & 50.00 & 100.00 & 33.33 & 57.69 & 65.96 & 23.99\tabularnewline
        Job/task-Assessment, investigation, testing, screening (e.g., holistic
        review) & 66.67 & 42.86 & 50.00 & 100.00 & 50.00 & 63.46 & 62.16 & 20.60\tabularnewline
        Organisation-National guidance & 16.67 & 71.43 & 50.00 & 100.00 & 16.67 & 71.15 & 54.32 & 33.22\tabularnewline
        Job/task-Care Planning & 50.00 & 42.86 & 50.00 & 100.00 & 33.33 & 32.69 & 51.48 & 24.96\tabularnewline
        Person Patient-Physical characteristics & 33.33 & 57.14 & 75.00 & 0.00 & 83.33 & 48.08 & 49.48 & 30.22\tabularnewline
        Technologies and Tools-issues & 33.33 & 42.86 & 25.00 & 100.00 & 16.67 & 28.85 & 41.12 & 30.13\tabularnewline
        Job/task-Situation awareness (e.g., loss of helicopter view) & 33.33 & 28.57 & 50.00 & 100.00 & 0.00 & 25.00 & 39.48 & 33.76\tabularnewline
        Person Staff-Slip or lapse (errors that tend to happen in routine
        tasks that people are doing without much conscious thought) & 33.33 & 42.86 & 50.00 & 0.00 & 33.33 & 42.31 & 33.64 & 17.66\tabularnewline
        Organisation-Language barrier & 50.00 & 0.00 & 0.00 & 100.00 & 16.67 & 0.00 & 27.78 & 40.37\tabularnewline
        Organisation-Documentation & 50.00 & 14.29 & 25.00 & 0.00 & 16.67 & 53.85 & 26.63 & 21.21\tabularnewline
        Job/task-Monitoring & 50.00 & 14.29 & 25.00 & 0.00 & 33.33 & 26.92 & 24.92 & 16.96\tabularnewline
        Organisation-Escalation/referral factor (including fresh eyes reviews) & 16.67 & 28.57 & 50.00 & 0.00 & 16.67 & 36.54 & 24.74 & 17.52\tabularnewline
        Job/task-Obstetric review & 33.33 & 14.29 & 25.00 & 0.00 & 33.33 & 34.62 & 23.43 & 13.83\tabularnewline
        Job/task-Dispensing, administering & 0.00 & 0.00 & 100.00 & 0.00 & 0.00 & 13.46 & 18.91 & 40.09\tabularnewline
        Person Patient-Psychological characteristics (e.g., stress, mental
        health) & 16.67 & 14.29 & 50.00 & 0.00 & 16.67 & 15.38 & 18.83 & 16.54\tabularnewline
        Internal Environment-Acuity (e.g., capacity of the maternity unit
        as a whole) & 33.33 & 0.00 & 25.00 & 0.00 & 16.67 & 11.54 & 14.42 & 13.41\tabularnewline
        Person Staff-Training and education (e.g., attendance at antenatal
        classes) & 16.67 & 0.00 & 50.00 & 0.00 & 0.00 & 15.38 & 13.68 & 19.45\tabularnewline
        Job/task-Risk assessment & 33.33 & 28.57 & 0.00 & 0.00 & 0.00 & 13.46 & 12.56 & 15.24\tabularnewline
        Internal Environment-Physical layout and Environment & 0.00 & 14.29 & 25.00 & 0.00 & 0.00 & 5.77 & 7.51 & 10.24\tabularnewline
        Person Staff-Decision error (errors in conscious judgments, decisions
        due to lack of knowledge and from misunderstanding of a situation) & 0.00 & 0.00 & 0.00 & 0.00 & 16.67 & 23.08 & 6.62 & 10.46\tabularnewline
        Technologies and Tools (e.g., CTG, use of screening tools)-Interpretation
        (e.g., CTG) & 0.00 & 14.29 & 0.00 & 0.00 & 0.00 & 21.15 & 5.91 & 9.40\tabularnewline
        External Environment-Geographical factor (e.g. Location of patient) & 16.67 & 0.00 & 0.00 & 0.00 & 0.00 & 1.92 & 3.10 & 6.69\tabularnewline
        \hline 
        \multicolumn{9}{l}{DNR: Data not received, WO: White Other, WB: White British, MB: Mixed
        background, SD: Standard Deviation.}\tabularnewline
    
           \multicolumn{9}{p{6in}}{Formula for computing the percentages: {Number of reports in which a concept $i$ appeared for ethnicity $e$} / {Total number of reports in ethnic group $e$}}\tabularnewline
        \hline 
\end{longtable}

\begin{table}[!h]
	\centering
	\caption{Statistics about the investigation reports provided by HSIB (Batch 1)}
	\label{SupplementaryTable 3}
 \small
	\begin{tabular}{rcp{1.3in}p{1.3in}}
		\hline 
		Ethnic group & No. of reports & No. of concepts across the reports & Average no. of concepts per report\tabularnewline
		\hline 
		Asian & 6 & 87 & 15\tabularnewline
		Black & 7 & 81 & 12\tabularnewline
		Data not received & 4 & 55 & 14\tabularnewline
		Mixed Background & 1 & 13 & 13\tabularnewline
		White British & 52 & 688 & 14\tabularnewline
		White Other & 6 & 46 & 8\tabularnewline
  \hline
		 Total: & 76 & 970 & Average: 13 \tabularnewline
		\hline 
	\end{tabular}
\end{table}


\begin{longtable}[c]{p{3.2in}p{0.3in}p{0.3in}p{0.3in}p{0.3in}p{0.3in}p{0.40in}|p{0.35in}}
\caption{File and concept frequencies. Each cell of the table shows the number of files (total number of concepts) across each ethnic group for Batch 1.}
\label{SupplementaryTable 4}\tabularnewline
\hline 
Concept & Asian & Black & DNR & MB & WO & WB & Sum \tabularnewline \hline
Organisation-Communication factor & 5(20) & 7(12) & 3(4) & 1(4) & 4(6) & 39(93) & 59(139)\tabularnewline
Job/task-assessment, investigation, testing, screening (e.g., holistic
review) & 4(7) & 3(5) & 2(3) & 1(1) & 3(4) & 33(73) & 46(93)\tabularnewline
External Environment-COVID & 4(5) & 4(8) & 1(1) & 1(1) & 4(6) & 42(71) & 56(92)\tabularnewline
Organisation-Teamworking & 5(9) & 5(11) & 2(8) & 1(1) & 2(3) & 30(58) & 45(90)\tabularnewline
Organisation-National guidance & 1(2) & 5(7) & 2(4) & 1(2) & 1(1) & 37(73) & 47(89)\tabularnewline
Person Patient-Physical characteristics & 2(2) & 4(6) & 3(4) & -(-) & 5(6) & 25(41) & 39(59)\tabularnewline
Organisation-Documentation & 3(3) & 1(1) & 1(1) & -(-) & 1(1) & 28(43) & 34(49)\tabularnewline
Person Staff-Slip or lapse (errors that tend to happen in routine
tasks that people are doing without much conscious thought) & 2(3) & 3(4) & 2(3) & -(-) & 2(4) & 22(30) & 31(44)\tabularnewline
Job/task-Care Planning & 3(5) & 3(6) & 2(6) & 1(1) & 2(2) & 17(20) & 28(40)\tabularnewline
Organisation-Escalation/referral factor (including fresh eyes reviews) & 1(3) & 2(4) & 2(2) & -(-) & 1(1) & 19(28) & 25(38)\tabularnewline
Job/task-Obstetric review & 2(2) & 1(1) & 1(1) & -(-) & 2(3) & 18(24) & 24(31)\tabularnewline
Job/task-Monitoring & 3(5) & 1(1) & 1(1) & -(-) & 2(2) & 14(21) & 21(30)\tabularnewline
Technologies and Tools-issues & 2(2) & 3(5) & 1(2) & 1(1) & 1(2) & 15(17) & 23(29)\tabularnewline
Job/task-Situation awareness (e.g., loss of helicopter view) & 2(4) & 2(2) & 2(2) & 1(1) & 0(-) & 13(18) & 20(27)\tabularnewline
Person Staff-Decision error (errors in conscious judgments, decisions
due to lack of knowledge and from misunderstanding of a situation) & -(-) & -(-) & -(-) & -(-) & 1(2) & 12(17) & 13(19)\tabularnewline
Technologies and Tools (e.g., CTG, use of screening tools)-Interpretation
(e.g., CTG) & -(-) & 1(3) & -(-) & -(-) & -(-) & 11(13) & 12(16)\tabularnewline
Internal Environment-Acuity (e.g., capacity of the maternity unit
as a whole) & 2(3) & -(-) & 1(2) & -(-) & 1(1) & 6(9) & 10(15)\tabularnewline
Person Patient-Psychological characteristics (e.g., stress, mental
health) & 1(1) & 1(1) & 2(3) & -(-) & 1(1) & 8(9) & 13(15)\tabularnewline
Job/task-Dispensing, administering & -(-) & -(-) & 4(4) & -(-) & -(-) & 7(9) & 11(13)\tabularnewline
Person Staff-Training and education (e.g., attendance at antenatal
classes) & 1(1) & -(-) & 2(3) & -(-) & -(-) & 8(9) & 11(13)\tabularnewline
Job/task-Risk assessment & 2(2) & 2(3) & -(-) & -(-) & -(-) & 7(7) & 11(12)\tabularnewline
Organisation-Language barrier & 3(7) & -(-) & -(-) & 1(1) & 1(1) & -(-) & 5(9)\tabularnewline
Internal Environment-Physical layout and Environment & -(-) & 1(1) & 1(1) & -(-) & -(-) & 3(4) & 5(6)\tabularnewline
External Environment-Geographical factor (e.g. Location of patient) & 1(1) & -(-) & -(-) & -(-) & -(-) & 1(1) & 2(2)\tabularnewline
\hline
Total & 49(87) & 49(81) & 35(55) & 9(13) & 34(46) & 415(688) & 591(970)\tabularnewline
\hline 
\multicolumn{8}{l}{\small{DNR: Data not received, MB: Mixed Background, WO: White Other, WB: White British}}\tabularnewline
\multicolumn{8}{l}{\small{- denotes that the concept did not appear in the particular ethnic group's reports.}}\tabularnewline
\hline
\end{longtable}

\begin{table}[!h]
	\centering
	\caption{Batch 1 reports and their associated outcomes across the ethnic groups. This information was only provided with Batch 1.}
	\label{SupplementaryTable 5}
		\begin{tabular}{lcccc|c}
			\hline 
			Ethnic group & TH & NND & MD & Stillbirth & Total\tabularnewline
   \hline
			Asian & 3 & 1 & 1 & 1 & 6\tabularnewline
			Black & 3 & - & 4 & - & 7\tabularnewline
			Data not received & - & - & 4 & - & 4\tabularnewline
			Mixed Background & 1 & - & - & - & 1\tabularnewline
			White British & 30 & 5 & 8 & 9 & 52\tabularnewline
			White Other & 2 & - & 2 & 2 & 6\tabularnewline
   \hline
			Total & 39 & 6 & 19 & 12 & 76\tabularnewline
			\hline 
			\multicolumn{6}{l}{Therapeutic Hypothermia: TH, Early Neonatal Death: NND, Mother's death: MD.}\tabularnewline
   \multicolumn{6}{l}{\small{- denotes that the outcome did not appear in the particular ethnic group's reports.}}\tabularnewline
			\hline 
		\end{tabular}
\end{table}

\begin{table}[!h]
\begin{longtable}[c]{ccc|cc|cc}
\caption{Performance evaluation for each test. The results of Test A and Test C can be compared because both tests were conducted on Batch 4. Test B was conducted on Batch 3 and hence cannot be directly compared to the results of Test A and Test C.}
\label{SupplementaryTable 6}\tabularnewline
\hline 
&\multicolumn{2}{c|}{Test with real data} & \multicolumn{4}{c}{Tests with synthetic data} \tabularnewline
\hline 
& \multicolumn{2}{c|}{Test B} & \multicolumn{2}{c|}{Test A} & \multicolumn{2}{c}{Test C}\tabularnewline 
 & Avg & SD & Avg & SD & Avg & SD\tabularnewline
\hline 
Precision & 0.87 & 0.34 & 1.00 & 0.00 & 1.00 & 0.00\tabularnewline
Recall & 0.93 & 0.18 & 0.60 & 0.23 & 0.67 & 0.15\tabularnewline
F-score & 0.96 & 0.10 & 0.79 & 0.08 & 0.81 & 0.08\tabularnewline
Misclassification & 0.19 & 0.35 & 0.40 & 0.23 & 0.33 & 0.15\tabularnewline
Accuracy & 0.81 & 0.35 & 0.60 & 0.23 & 0.67 & 0.15\tabularnewline
Balanced Accuracy & 0.90 & 0.18 & 0.80 & 0.11 & 0.83 & 0.08\tabularnewline
\hline 
\multicolumn{7}{l}{Avg: Average; SD: Standard deviation. }\tabularnewline
\hline 
\end{longtable}
\end{table}

\begin{longtable}[!c]{p{10cm}c|cc}
\caption{Percentage of correctly annotated concepts across all the tests A, B, and C}
\label{SupplementaryTable 7}
        \tabularnewline
        &\multicolumn{1}{c}{Test with real data} &\multicolumn{2}{c}{Tests with synthetic data} \tabularnewline
        \hline 
        Concept & Test B & Test A & Test C\tabularnewline
        \hline 
        Organisation-Communication factor & 82.70 & 45.59 & 62.32\tabularnewline
        Job/task-Assessment, investigation, testing, screening (e.g., holistic
        review) & 93.39 & 60.97 & 61.26\tabularnewline
        External Environment-COVID & 86.67 & 55.59 & 59.95\tabularnewline
        Organisation-Teamworking & 90.53 & 74.40 & 60.58\tabularnewline
        Organisation-National guidance & 94.02 & 78.22 & 73.25\tabularnewline
        Person Patient-Physical characteristics & 90.42 & 66.81 & 71.98\tabularnewline
        Organisation-Documentation & 86.73 & 79.05 & 79.59\tabularnewline
        Person Staff-Slip or lapse (errors that tend to happen in routine
        tasks that people are doing without much conscious thought) & 91.27 & 80.00 & 87.78\tabularnewline
        Job/task-Care Planning & 93.90 & 80.74 & 81.27\tabularnewline
        Organisation-Escalation/referral factor (including fresh eyes reviews) & 92.73 & 61.86 & 57.10\tabularnewline
        Job/task-Obstetric review & 95.10 & 48.36 & 52.36\tabularnewline
        Job/task-Monitoring & 97.33 & 46.38 & 63.65\tabularnewline
        Technologies and Tools-issues & 78.79 & 80.85 & 79.97\tabularnewline
        Job/task-Situation awareness (e.g., loss of helicopter view) & 75.00 & 49.33 & 57.00\tabularnewline
        Person Staff-Decision error (errors in conscious judgements, decisions
        due to lack of knowledge and from misunderstanding of a situation) & 92.68 & 37.50 & 35.00\tabularnewline
        Technologies and Tools (e.g., CTG, use of screening tools)-Interpretation
        (e.g., CTG) & 100.00 & 51.93 & 27.88\tabularnewline
        Internal Environment-Acuity (e.g., capacity of the maternity unit
        as a whole) & 83.33 & 12.50 & 11.43\tabularnewline
        Person Patient-Psychological characteristics (e.g., stress, mental
        health) & 72.73 & 52.73 & 52.73\tabularnewline
        Job/task-Dispensing, administering & 92.86 & 36.37 & 36.36\tabularnewline
        Person Staff-Training and education (e.g., attendance at antenatal
        classes) & 77.27 & 96.97 & 90.91\tabularnewline
        Job/task-Risk assessment & 98.61 & 58.33 & 72.50\tabularnewline
        Organisation-Language barrier & - & 0.00 & 22.22\tabularnewline
        Internal Environment-Physical layout and Environment & 82.61 & 53.33 & 53.33\tabularnewline
        External Environment-Geographical factor (e.g. Location of patient) & - & 50.00 & 50.00\tabularnewline
        Organisation-Guidance factor & 100.00 & - & -\tabularnewline
        Person Patient-Language barrier & 95.24 & - & -\tabularnewline
        Organisation-Local guidance & 92.11 & - & -\tabularnewline
        Organisation-National and local guidance & 98.00 & - & -\tabularnewline
        \hline 
        Average & 89.77 & 56.57 & 58.35\tabularnewline
        Standard Deviation & 7.85 & 22.02 & 20.50\tabularnewline
        \multicolumn{4}{l}{- denotes that the concept did not appear in the particular Test.}\tabularnewline \hline
\end{longtable}

\begin{longtable}{p{7cm}cccccc|cc}
 \caption{Percentage of correctly annotated concepts in the synthetic (test) sentences for each ethnic group. Trained on real-data (Batch 1) and tested on synthetic data (Batch 4). These are the results from testing with Batch 4 (Test A).}
	\label{SupplementaryTable 8}\tabularnewline
        \hline 
        Code & Asian & Black & DNR & MB & OW & WB & Avg & SD\tabularnewline
        \hline 
        External Environment-COVID & 64.29 & 83.33 & 50.00 & 0.00 & 66.67 & 69.23 & 55.59 & 29.24\tabularnewline
        External Environment-Geographical factor (e.g. Location of patient) & 0.00 & - & - & - & - & 100.00 & 50.00 & 70.71\tabularnewline
        Internal Environment-Acuity (e.g., capacity of the maternity unit
        as a whole) & 0.00 & - & 0.00 & - & 0.00 & 50.00 & 12.50 & 25.00\tabularnewline
        Internal Environment-Physical layout and Environment & - & 0.00 & 100.00 & - & - & 60.00 & 53.33 & 50.33\tabularnewline
        Job/task-Assessment, investigation, testing, screening (e.g., holistic
        review) & 57.14 & 75.00 & 80.00 & 0.00 & 80.00 & 73.68 & 60.97 & 31.03\tabularnewline
        Job/task-Care Planning & 100.00 & 66.67 & 100.00 & 100.00 & 50.00 & 67.74 & 80.74 & 22.02\tabularnewline
        Job/task-Dispensing, administering & - & - & 0.00 & - & - & 72.73 & 36.37 & 51.43\tabularnewline
        Job/task-Monitoring & 66.67 & 0.00 & 0.00 & - & 100.00 & 65.22 & 46.38 & 44.56\tabularnewline
        Job/task-Obstetric review & 0.00 & 50.00 & 50.00 & - & 60.00 & 81.82 & 48.36 & 30.00\tabularnewline
        Job/task-Risk assessment & 0.00 & 100.00 & - & - & - & 75.00 & 58.33 & 52.04\tabularnewline
        Job/task-Situation awareness (e.g., loss of helicopter view) & 100.00 & 33.33 & 33.33 & 0.00 & - & 80.00 & 49.33 & 40.17\tabularnewline
        Organisation-Communication factor & 40.00 & 50.00 & 50.00 & 0.00 & 60.00 & 73.53 & 45.59 & 25.05\tabularnewline
        Organisation-Documentation & 75.00 & 100.00 & 50.00 & - & 100.00 & 70.27 & 79.05 & 21.30\tabularnewline
        Organisation-Escalation/referral factor (including fresh eyes reviews) & 60.00 & 66.67 & 100.00 & - & 0.00 & 82.61 & 61.86 & 37.88\tabularnewline
        Organisation-Language barrier & 0.00 & - & - & 0.00 & 0.00 & - & 0.00 & 0.00\tabularnewline
        Organisation-National guidance & 83.33 & 62.50 & 66.67 & 100.00 & 80.00 & 76.79 & 78.22 & 13.31\tabularnewline
        Organisation-Teamworking & 66.67 & 66.67 & 53.85 & 100.00 & 83.33 & 75.86 & 74.40 & 15.99\tabularnewline
        Person Patient-Physical characteristics & 100.00 & 66.67 & 0.00 & - & 100.00 & 67.39 & 66.81 & 40.83\tabularnewline
        Person Patient-Psychological characteristics (e.g., stress, mental
        health) & 100.00 & 100.00 & 0.00 & - & 0.00 & 63.64 & 52.73 & 50.37\tabularnewline
        Person Staff-Decision error (errors in conscious judgements, decisions
        due to lack of knowledge and from misunderstanding of a situation) & - & - & - & - & 0.00 & 75.00 & 37.50 & 53.03\tabularnewline
        Person Staff-Slip or lapse (errors that tend to happen in routine
        tasks that people are doing without much conscious thought) & 100.00 & 80.00 & 50.00 & - & 100.00 & 70.00 & 80.00 & 21.21\tabularnewline
        Person Staff-Training and education (e.g., attendance at antenatal
        classes) & 100.00 & - & 100.00 & - & - & 90.91 & 96.97 & 5.25\tabularnewline
        Technologies and Tools (e.g., CTG, use of screening tools)-Interpretation
        (e.g., CTG) & - & 50.00 & - & - & - & 53.85 & 51.93 & 2.72\tabularnewline
        Technologies and Tools-issues & 100.00 & 100.00 & 66.67 & 100.00 & 50.00 & 68.42 & 80.85 & 21.94\tabularnewline
 \hline
        \multicolumn{9}{l}{DNR: Data not received, WO: White Other, WB: White British, MB: Mixed
        background, SD: Standard Deviation, Avg: Average}\tabularnewline
        \multicolumn{9}{l}{- denotes that the concept did not appear in the particular ethnic group.}\tabularnewline
        \hline
	\end{longtable}
\newpage
\centering
        \begin{longtable}{l|ccc|ccc}
        \caption{Test performance when training on Batch 1 and testing on Batch 4.
Before retraining (Test A): Training on Batch 1 and testing on Batch 4. After retraining (Test C): Retraining on Batches 2 and 3 and testing on Batch 4.}
	\label{SupplementaryTable 9}\tabularnewline
        \hline 
        \multicolumn{1}{l}{} & \multicolumn{3}{c|}{Test A} & \multicolumn{3}{c}{Test C}\tabularnewline
        \hline 
        Ethnicity & Correctly & Incorrectly & Correct (\%) & Correctly & Incorrectly & Correct(\%)\tabularnewline
        \hline 
        Asian & 59 & 28 & 67.82 & 61 & 26 & 70.11\tabularnewline
        Black & 54 & 27 & 66.67 & 54 & 27 & 66.67\tabularnewline
        Data not received & 33 & 22 & 60.00 & 32 & 23 & 58.18\tabularnewline
        Mixed Background & 9 & 4 & 69.23 & 9 & 4 & 69.23\tabularnewline
        Other White & 32 & 14 & 69.57 & 34 & 12 & 73.91\tabularnewline
        White British & 499 & 189 & 72.82 & 501 & 187 & 72.82\tabularnewline
        \hline
        & Sum: 686 & Sum: 284 & Mean: 70.93($\pm$ 4.30) & Sum: 691 & Sum: 279 & Mean: 71.24 ($\pm{}$ 5.67)\tabularnewline
        \hline 
        \end{longtable}

\begin{longtable}{p{7cm}cccccc|cc}
 \caption{Percentage of correctly annotated concepts in synthetic sentences for each ethnic group (after retraining with Batch 3). Trained with Batch 1, retrained with Batches 2 and 3, and tested on Batch 4. These are the results from testing with Batch 4 (Test C).}
	\label{SupplementaryTable 10}\tabularnewline
\hline 
Concepts & Asian & Black & DNR & MB & OW & WB & Avg & SD\tabularnewline
        \hline 
        External Environment-COVID & 57.14 & 83.33 & 50.00 & 0.00 & 100.00 & 69.23 & 59.95 & 34.46\tabularnewline
        External Environment-Geographical factor (e.g. Location of patient) & 0.00 & - & - & - & - & 100.00 & 50.00 & 70.71\tabularnewline
        Internal Environment-Acuity (e.g., capacity of the maternity unit
        as a whole) & 0.00 & 0.00 & 0.00 & - & 0.00 & 57.14 & 11.43 & 25.56\tabularnewline
        Internal Environment-Physical layout and Environment & - & 0.00 & 100.00 & - & - & 60.00 & 53.33 & 50.33\tabularnewline
        Job/task-Assessment, investigation, testing, screening (e.g., holistic
        review) & 57.14 & 75.00 & 80.00 & 0.00 & 80.00 & 75.44 & 61.26 & 31.18\tabularnewline
        Job/task-Care Planning & 100.00 & 66.67 & 100.00 & 100.00 & 50.00 & 70.97 & 81.27 & 21.68\tabularnewline
        Job/task-Dispensing, administering & - & - & 0.00 & - & - & 72.73 & 36.36 & 51.43\tabularnewline
        Job/task-Monitoring & 66.67 & 50.00 & 0.00 & 100.00 & 100.00 & 65.22 & 63.65 & 37.13\tabularnewline
        Job/task-Obstetric review & 0.00 & 50.00 & 50.00 & - & 80.00 & 81.82 & 52.36 & 33.11\tabularnewline
        Job/task-Risk assessment & 0.00 & 100.00 & 100.00 & - & 100.00 & 62.50 & 72.50 & 43.66\tabularnewline
        Job/task-Situation awareness (e.g., loss of helicopter view) & 100.00 & 66.67 & 33.33 & 0.00 & - & 85.00 & 57.00 & 40.42\tabularnewline
        Organisation-Communication factor & 53.33 & 40.00 & 50.00 & 100.00 & 60.00 & 70.59 & 62.32 & 21.09\tabularnewline
        Organisation-Documentation & 100.00 & 75.00 & 50.00 & - & 100.00 & 72.97 & 79.59 & 21.06\tabularnewline
        Organisation-Escalation/referral factor (including fresh eyes reviews) & 60.00 & 50.00 & 50.00 & 100.00 & 0.00 & 82.61 & 57.10 & 34.24\tabularnewline
        Organisation-Language barrier & 0.00 & - & - & 0.00 & 0.00 & 88.89 & 22.22 & 44.44\tabularnewline
        Organisation-National guidance & 83.33 & 62.50 & 33.33 & 100.00 & 80.00 & 80.36 & 73.25 & 22.90\tabularnewline
        Organisation-Teamworking & 66.67 & 66.67 & 69.23 & 0.00 & 83.33 & 77.59 & 60.58 & 30.42\tabularnewline
        Person Patient-Physical characteristics & 100.00 & 66.67 & 0.00 & 100.00 & 100.00 & 65.22 & 71.98 & 39.01\tabularnewline
        Person Patient-Psychological characteristics (e.g., stress, mental
        health) & 100.00 & 100.00 & 0.00 & - & 0.00 & 63.64 & 52.73 & 50.37\tabularnewline
        Person Staff-Decision error (errors in conscious judgements, decisions
        due to lack of knowledge and from misunderstanding of a situation) & 0.00 & - & 100.00 & 0.00 & 0.00 & 75.00 & 35.00 & 48.73\tabularnewline
        Person Staff-Slip or lapse (errors that tend to happen in routine
        tasks that people are doing without much conscious thought) & 100.00 & 100.00 & 50.00 & 100.00 & 100.00 & 76.67 & 87.78 & 20.73\tabularnewline
        Person Staff-Training and education (e.g., attendance at antenatal
        classes) & 100.00 & - & 100.00 & - & - & 72.73 & 90.91 & 15.75\tabularnewline
        Technologies and Tools (e.g., CTG, use of screening tools)-Interpretation
        (e.g., CTG) & - & 50.00 & 0.00 & - & 0.00 & 61.54 & 27.88 & 32.54\tabularnewline
        Technologies and Tools-issues & 100.00 & 100.00 & 66.67 & 100.00 & 50.00 & 63.16 & 79.97 & 22.63\tabularnewline
        \hline 
        \multicolumn{9}{l}{DNR: Data not received, WO: White Other, WB: White British, MB: Mixed
        background, SD: Standard Deviation, Avg: Average}\tabularnewline
    
          \multicolumn{9}{l}{- denotes that the concept did not appear in the particular ethnic group.}\tabularnewline
         
\end{longtable}
\vspace{-0.1cm}
    \begin{longtable}{lcc}
    \caption{Wilcoxon signed-rank test results: Comparing Tests A and C.}
    \label{SupplementaryTable 11}\tabularnewline
    \hline 
    Ethnicity & W & p\tabularnewline
    \hline 
    Asian & 1 & 0.29\tabularnewline
    Black & 7 & 0.46\tabularnewline
    Data not received & 1 & 0.29\tabularnewline
    Mixed Background & 1.5 & 1.00\tabularnewline
    Other White & 0 & 0.18\tabularnewline
    White British & 44 & 0.59\tabularnewline
    \hline 
    \end{longtable}

    
\begin{longtable}{cp{6cm}p{2cm}p{2.5cm}p{2.5cm}}
    \caption{Sample text segments annotations by Human Annotators using the SIRch framework.}
    \label{SupplementaryTable 12} \tabularnewline
        \hline
        \multicolumn{1}{c}{\textbf{Report ID}} & \multicolumn{1}{c}{\textbf{Sentence}} & \multicolumn{1}{c}{\textbf{HA-1}} & \multicolumn{1}{c}{\textbf{HA-2}} & \multicolumn{1}{c}{\textbf{HA-3}}\tabularnewline
        \hline
        ID-1 & This meant that during pregnancy the Mother did not have an \textcolor{orange}{anaesthetic review}, \textcolor{teal}{referral} to a dietitian, or \textcolor{blue}{information shared} regarding the risks associated with raised BMI during pregnancy and the need for high dose folic acid. & \textcolor{orange}{Assessment, investigation, testing, screening (e.g., holistic review)} & \textcolor{teal}{Escalation/referral factor (including fresh eyes reviews)}   & \textcolor{blue}{Between staff and patient (verbal)}\tabularnewline
        ID-2 & The \textcolor{orange}{abnormal CTG} was not recognised and did not prompt \textcolor{teal}{escalation to the obstetric team}, in line with \textcolor{blue}{local guidance}. & \textcolor{orange}{Interpretation (e.g., CTG)} & \textcolor{teal}{Escalation/ referral factor (including fresh eyes reviews)}   & \textcolor{blue}{Local guidance}\tabularnewline
        ID-3 & \textcolor{orange}{Obstetric reviews} of the Mother whilst in labour were not carried out in line with \textcolor{teal}{local guidance}, these would have provided an earlier opportunity for a \textcolor{blue}{holistic review} of the Mother's risk status and care pathway. & \textcolor{orange}{Obstetric review}   & \textcolor{teal}{Local guidance} & \textcolor{blue}{Assessment, investigation, testing, screening (e.g., holistic review)}\tabularnewline
        \hline
        \multicolumn{5}{l}{HA: Human Annotator}\tabularnewline
        \hline
\end{longtable}
\newpage
\begin{longtable}[!c]{p{10cm}ccc|c}
\caption{Number of files and concept frequencies found in the reports that were utilized for developing the model. Batch information is provided in Supp. Table S\ref{SupplementaryTable 1}. The columns show the Number of files (concept frequency).}
\label{SupplementaryTable 13}
        \tabularnewline
        \hline 
        Concepts & Batch 1 & Batch 2 & Batch 3 & Total\tabularnewline
        \hline 
        Organisation-Teamworking & 45(90) & 14(81) & 96(359) & 155(530)\tabularnewline
        Organisation-Communication factor & 59(139) & 15(48) & 85(290) & 159(477)\tabularnewline
        Job/task-Assessment, investigation, testing, screening (e.g., holistic
        review) & 46(93) & 14(46) & 90(242) & 150(381)\tabularnewline
        Person Patient-Physical characteristics & 39(59) & 10(22) & 69(240) & 118(321)\tabularnewline
        Technologies and Tools (e.g., CTG, use of screening tools)-Interpretation
        (e.g., CTG) & 12(16) & 13(35) & 65(146) & 90(197)\tabularnewline
        Person Staff-Slip or lapse (errors that tend to happen in routine
        tasks that people are doing without much conscious thought) & 31(44) & 7(18) & 61(126) & 99(188)\tabularnewline
        Person Staff-Decision error (errors in conscious judgements, decisions
        due to lack of knowledge and from misunderstanding of a situation) & 13(19) & 12(27) & 64(123) & 89(169)\tabularnewline
        Organisation-Documentation & 34(49) & 7(21) & 45(98) & 86(168)\tabularnewline
        Organisation-Escalation/referral factor (including fresh eyes reviews) & 25(38) & 8(10) & 64(110) & 97(158)\tabularnewline
        Organisation-National and local guidance & 47(89) & 8(11) & 37(50) & 92(150)\tabularnewline
        Job/task-Obstetric review & 24(31) & 7(14) & 58(102) & 89(147)\tabularnewline
        External Environment-COVID & 56(92) & 2(5) & 21(45) & 79(142)\tabularnewline
        Organisation-National guidance & 0(0) & 11(18) & 76(117) & 87(135)\tabularnewline
        Job/task-Care Planning & 28(40) & 6(10) & 49(82) & 83(132)\tabularnewline
        Job/task-Monitoring & 21(30) & 8(13) & 47(75) & 76(118)\tabularnewline
        Technologies and Tools-issues & 23(29) & 9(18) & 37(66) & 69(113)\tabularnewline
        Job/task-Risk assessment & 11(12) & 4(10) & 46(72) & 61(94)\tabularnewline
        Job/task-Situation awareness (e.g., loss of helicopter view) & 20(27) & 1(2) & 30(48) & 51(77)\tabularnewline
        Job/task-Dispensing, administering & 11(13) & 4(7) & 25(42) & 40(62)\tabularnewline
        Internal Environment-Acuity (e.g., capacity of the maternity unit
        as a whole) & 10(15) & 3(4) & 32(36) & 45(55)\tabularnewline
        Organisation-Local guidance & 0(0) & 9(17) & 27(38) & 36(55)\tabularnewline
        Person Patient-Psychological characteristics (e.g., stress, mental
        health) & 13(15) & 3(5) & 24(33) & 40(53)\tabularnewline
        Person Staff-Training and education (e.g., attendance at antenatal
        classes) & 11(13) & 2(5) & 12(22) & 25(40)\tabularnewline
        Organisation-Guidance factor & 0(0) & 1(1) & 26(38) & 27(39)\tabularnewline
        Internal Environment-Physical layout and Environment & 5(6) & 3(4) & 13(23) & 21(33)\tabularnewline
        Person Patient-Language barrier & 5(9) & 0(0) & 10(21) & 15(30)\tabularnewline
        External Environment-Geographical factor (e.g. Location of patient) & 2(2) & 0(0) & 0(0) & 2(2)\tabularnewline
        \hline 
        Total & 591(970) & 181(452) & 1209(2644) & 1981(4066)\tabularnewline
        \hline 
\end{longtable}

\begin{table}
    \centering
    \caption{Inter-rater reliability among three human annotators (HA). The columns show the number of sentences per report coded by each HA. The last column shows the number of sentences for which there was agreement on the annotations.}
    \label{SupplementaryTable 14}
    \begin{tabular}{cccc|p{1in}}
        \hline
            RN. & HA-1 & HA-2 & HA-3 & Agreement \tabularnewline
            \hline 
            1 & 6 & 6 & 6 & 5 \tabularnewline
            2 & 8 & 8 & 8 & 6 \tabularnewline
            3 & 6 & 6 & 6 & 5 \tabularnewline
            4 & 10 & 10 & 11 & 8\tabularnewline
            5 & 5 & 5 & 5 & 5 \tabularnewline
            6 & 5 & 5 & 5 & 5 \tabularnewline
            7 & 20 & 20 & 20 & 16\tabularnewline
            8 & 10 & 10 & 7 & 7 \tabularnewline
            9 & 8 & 8 & 8 & 7 \tabularnewline
            10 & 6 & 6 & 6 & 5 \tabularnewline
            11 & 29 & 29 & 25 & 17 \tabularnewline
            12 & 7 & 7 & 6 & 5 \tabularnewline
            13 & 7 & 7 & 11 & 6 \tabularnewline
            14 & 6 & 6 & 5 & 5 \tabularnewline
            15 & 5 & 5 & 5 & 5 \tabularnewline
            16 & 5 & 5 & 4 & 4 \tabularnewline
            17 & 10 & 10 & 8 & 6 \tabularnewline
            18 & 13 & 13 & 15 & 14 \tabularnewline
            19 & 5 & 5 & 5 & 5 \tabularnewline
            20 & 5 & 5 & 6 & 4 \tabularnewline
            \hline 
            \multicolumn{5}{p{3in}}{RN.: Report number, HA: Human Annotator}  \tabularnewline
            \hline 
            \multicolumn{5}{p{3in}}{Total number of concept agreements: 420; Total number of concepts: 524; Number of annotators: 3. 
            IRR: 80.15\%} \tabularnewline
            \hline 
               \end{tabular}
        \end{table}

\begin{table}
\begin{longtable}{lp{3in}p{3in}}
    \caption{Sample of original and synthetic (test) sentences. In the dataset only Batch 4 contained synthetic sentences, and these were only used for testing purposes and not for training the model.}
\label{SupplementaryTable 15}\tabularnewline
        \hline
      No. & Original sentences & Synthetic sentences\tabularnewline
        \hline 
        1 & The Mother was appropriately risk assessed at booking for midwifery
        led care and the pregnancy progressed without complication until the
        induction of her labour at 41+5 weeks which was offered in line with
        national guidance. & The Mother's risk was assessed when she booked for midwifery led care
        and her pregnancy progressed without complication until she was induced
        at 41+5 weeks, which was offered in line with national guidance.\tabularnewline
        2 & This would not have altered the outcome for the Baby as the Baby's
        birthweight was within the expected range. & Even if the outcome had been different, the baby's birthweight was
        still within the expected range.\tabularnewline
        3 & This had an impact on the Mother's experience as she did not feel
        able to have in depth discussions over the telephone. & This impacted the Mother's experience as she felt unable to have in
        depth discussions over the telephone.\tabularnewline
        4 & The Trust are encouraged to improve cross boundary working practices
        to ensure all mothers have equal access to local information about
        their care pathways. & The Trust should work on improving communication and collaboration
        between different departments to make sure all mothers have the same
        access to information about their care options.\tabularnewline
        5 & The Mother was given the opportunity for a discussion about her choices
        and IOL processes on admission to the maternity unit. & The Mother was given the opportunity to discuss her choices and IOL
        processes with a staff member upon admission to the maternity unit.\tabularnewline
        \hline
\end{longtable}
\end{table}

\newpage
\newpage
\section*{Supplementary File S{F1}. SIRch taxonomy of human factors for maternity investigations}
\label{sec:SIRchFramework}
\vspace{-.1cm}
\begin{mdframed}
\begin{multicols}{2}
\begin{enumerate}[itemsep=0pt,parsep=0pt]
\item External Environment
  \begin{enumerate}[itemsep=0pt,parsep=0pt]
  \item Policy factor 
  \item Societal factor
  \item Economic factor
  \item COVID \checkmark
  \item Geographical factor (e.g. Location of patient)
  \end{enumerate}
\item Internal Environment
  \begin{enumerate}[itemsep=0pt,parsep=0pt]
  \item Physical layout and Environment
  \item Acuity (e.g., capacity of the maternity unit as a whole)
  \item Availability (e.g., operating theatres)
  \item Time of day (e.g., night working or day of the week)
  \end{enumerate}
\item Organisation
  \begin{enumerate}[itemsep=0pt,parsep=0pt]
  \item Team culture factor (e.g., patient safety culture)
  \item Incentive factor (e.g., performance evaluation)
  \item Teamworking
  \item Communication factor
  \begin{enumerate}[itemsep=0pt,parsep=0pt]
      \item Between staff 
      \item Between staff and patient (verbal)
  \end{enumerate}
  \item Documentation
  \item Escalation/referral factor (including fresh eyes reviews)
  \item National and/or local guidance
  \item Language barrier
  \end{enumerate}
\item Jobs/Task
  \begin{enumerate}[itemsep=0pt,parsep=0pt]
  \item Assessment, investigation, testing, screening (e.g., holistic review) 
  \item Care planning 
  \item Dispensing, administering
  \item Monitoring 
  \item Risk assessment
  \item Situation awareness (e.g., loss of helicopter view)
  \item Obstetric review
  \end{enumerate}
\item Technologies and Tools
  \begin{enumerate}[itemsep=0pt,parsep=0pt]
  \item Issues
  \item Interpretation (e.g., CTG) 
  \end{enumerate}
\item Person
  \begin{enumerate}[itemsep=0pt,parsep=0pt]
  \item Patient (characteristics and performance)
  \begin{enumerate}[itemsep=0pt,parsep=0pt]
      \item Characteristics
      \begin{enumerate}[itemsep=0pt,parsep=0pt]
          \item Physical characteristics
          \item Psychological characteristics (e.g., stress, mental health)
          \item Language competence (English)
          \item Disability (e.g., hearing problems)
          \item Training and education (e.g., attendance at ante-natal classes)
          \item Record of attendance (e.g., failure to attend antenatal classes)
      \end{enumerate}
      \item Performance
      \begin{enumerate}[itemsep=0pt,parsep=0pt]
          \item Slip or lapse (errors that tend to happen in routine tasks that people are doing without much conscious thought)
          \item Decision error (errors in conscious judgements, decisions due to lack of knowledge and from misunderstanding of a situation)
          \item Intentional rule breaking (deliberately do something different from rules)
      \end{enumerate}
  \end{enumerate}
  \item Staff (characteristics and performance)
  \begin{enumerate}[itemsep=0pt,parsep=0pt]
      \item Characteristics
      \begin{enumerate}[itemsep=0pt,parsep=0pt]
          \item Physical characteristics 
          \item Psychological characteristics (e.g., stress, mental health)
          \item Language competence (English)
          \item Disability (e.g., hearing problems)
          \item Training and education (e.g., attendance at ante-natal classes)
          \item Record of attendance (e.g., failure to attend antenatal classes)     
      \end{enumerate}
      \item Performance
      \begin{enumerate}[itemsep=0pt,parsep=0pt]
          \item Slip or lapse (errors that tend to happen in routine tasks that people are doing without much conscious thought)
          \item Decision error (errors in conscious judgements, decisions due to lack of knowledge and from misunderstanding of a situation)
          \item Intentional rule breaking (deliberately do something different from rules)
      \end{enumerate}
  \end{enumerate}
  \end{enumerate}
\end{enumerate}
\end{multicols}
\end{mdframed}

\begin{figure}[h!] 
\centering
    \includegraphics[width=0.6\textwidth]{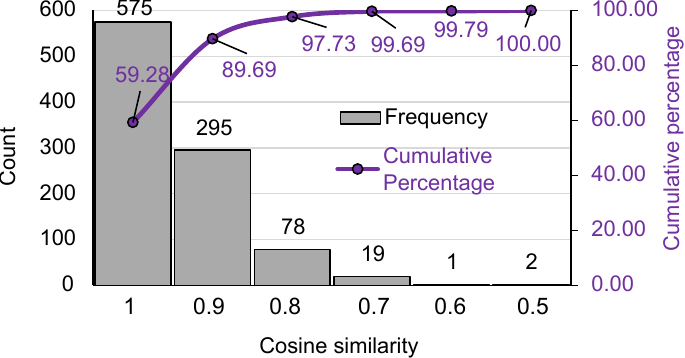}
\captionsetup{name=Figure,labelsep=period}
    \renewcommand{\thefigure}{S1}
    \caption{Distribution of cosine similarity values (semantic similarity). The higher the similarity value the closer the semantic meaning of the original and synthetic sentence (Batch 1).}
    \label{SupplementaryFigure1}
\end{figure}

\newpage
\begin{justify}
\section*{Supplementary File S{F2}. Synthetic Dataset Generation for testing purposes} 
\label{sec:syntheticdataset}

\paragraph{\textbf{Generate synthetic sentences.}} To test the performance of the model on new sentences, a text generation tool was implemented that takes as input a set of original sentences from the `Extracted text segments' and generates a new synthetic sentence per original sentence. The original and generated sentences were exported to a CSV file, with each row containing one original and its corresponding synthetic sentence. 
The synthetic sentences are then utilized for two purposes: 1) to further train the model by increasing its exposure to linguistic variability during training; and 2) to evaluate the performance of the trained model on new content that is semantically similar but uses different sentences. Supp. Table S\ref{SupplementaryTable 15} shows examples of original and synthetic sentences. 

\paragraph{\textbf{Sentence Similarity.}} It is important to understand how semantically similar the original sentences are to their corresponding synthetic sentences to evaluate the quality of the sentences generated by the synthetic sentence generation tool. To compute the similarity of each pair (i.e. original and synthetic sentence) a Word2Vec model was utilized, sentences were represented as numerical vectors based on their constituent words, capturing contextual meanings. The average vector for each sentence was then calculated by combining its word vectors. A mathematical measure, cosine similarity, was employed to compare these average sentence vectors, quantifying their semantic similarity or difference. A high cosine similarity score indicates strong semantic resemblance, while a low score suggests dissimilarity. In essence, this process allowed for the assessment of the likeness or disparity in meaning between two sentences using their word representations and cosine similarity. The similarity values of each pair were written to a CSV file. A statistical overview of the similarity values is given in Figure \ref{SupplementaryFigure1}, which shows that more than 97.5\% of the sentences have a cosine similarity of more than 80\% indicating good preservation of semantic content by the synthetic data generation tool. Hence, the tool effectively generated unique sentences while preserving the core meaning of the original sentence.
\end{justify}

\end{document}